\documentclass[%
 reprint,
 amsmath,amssymb,
 aps,
]{revtex4-2}

\usepackage{graphicx}
\usepackage{dcolumn}
\usepackage{bm}
\usepackage{xcolor}

\begin{document}

\preprint{APS/123-QED}

\title{Thermostatting of Active Hamiltonian Systems via Symplectic Algorithms}

\author{Antik Bhattacharya$^{1}$}
\email{abhattacharya@tifrh.res.in}
\author{J\"{u}rgen Horbach$^{2}$} 
\email{horbach@thphy.uni-duesseldorf.de}
\author{Smarajit Karmakar$^{1}$}
\email{smarajit@tifrh.res.in}

\affiliation{
$^{1}$Tata Institute of Fundamental Research Hyderabad, 36/P, Gopanpally Village, Serilingampally Mandal, Ranga Reddy District, Hyderabad, Telangana 500046, India}

\affiliation{
$^{2}$Institut f\"ur Theoretische Physik II: Weiche Materie, Heinrich-Heine-Universit\"at D\"usseldorf, Universit\"atsstraße 1, 40225 D\"usseldorf, Germany}

\date{\today}

\begin{abstract}
We consider a class of non-standard, two-dimensional (2D) Hamiltonian models that may show features of active particle dynamics, and therefore, we refer to these models as active Hamiltonian (AH) systems. The idea is to consider a spin fluid where -- on top of spin-spin and particle-particle interactions -- spins are coupled to the particle's velocities via a vector potential. Continuous spin variables interact with each other as in a standard $XY$ model. Typically, the AH models exhibit non-standard thermodynamic properties (e.g.~for temperature and pressure) and equations of motion with non-standard forces. This implies that the derivation of symplectic algorithms to solve Hamilton's equations of motion numerically, as well as the thermostatting for these systems, is not straightforward. Here, we derive a symplectic integration scheme and propose a Nos\'e-Poincar\'e thermostat, providing a correct sampling in the canonical ensemble. The expressions for AH systems that we find for temperature and pressure might have parallels with the ongoing debate about the definition of pressure and the equation of state in active matter systems. For a specific AH model, recently proposed by Casiulis {\it et al.}~[Phys. Rev. Lett. {\bf 124}, 198001 (2020)], we rationalize the symplectic algorithm and the proposed thermostatting, and investigate the transition from a fluid at high temperature to a cluster phase at low temperature where, due to the coupling of velocities and spins, the cluster phase shows a collective motion that is reminiscent to that observed in a variety of active systems.
\end{abstract}

\maketitle
\section{Introduction}
Active matter refers to a wide class of living and non-living systems \cite{vicsek1995novel, toner1998flocks, ramaswamy2010mechanics, palacci2013living, marchetti2013hydrodynamics} that contain particles or other units that can consume energy from the surroundings and transform it into mechanical energy. Examples range from flocks of birds and molecular motors to self-propelling colloidal particles. Active systems are inherently out-of-equilibrium and do not follow detailed balance. As such, they show a large variety of fascinating and complex dynamical phenomena. Among them is the formation of ordered phases of clusters or flocks \cite{vicsek1995novel, toner1998flocks} that undergo coherent collective motion at low noise strength and high particle density. Furthermore, many biological systems exhibit collective dynamical behavior in which forces generated by ATP consumption drive the dynamics instead of thermal fluctuations. A simple model of these systems that can capture some of the salient dynamical behaviours is a collection of self-propelled particles \cite{marchetti2013hydrodynamics}. It seems that typical patterns of active dynamics, such as flocking, do not have any counterpart in thermal equilibrium systems. At best, one might be able to mimic them in passive systems by applying (complicated) external fields that couple to the active entities.

Recently, however, Casiulis {\it et al.}~\cite{Casiulis2020_1,Casiulis2020_2} have proposed a two-dimensional Hamiltonian model that exhibits dynamical features such as the motion of moving ordered clusters of particles with a non-zero center-of-mass velocity. In this model, in the following referred to as CTCD model, $N$ particles interact with each other by a short-range pair potential, and in addition, the particles carry ferromagnetically-coupled continuous spins that are locally coupled to their own velocities via a simple scalar product. The CTCD model could be extended to other more general spin-velocity couplings, and thus, this approach opens the door towards studying dynamical patterns based on equilibrium Hamiltonian dynamics that show similarities to active particle systems. In the following, we refer to such systems as active Hamiltonian (AH) systems. Unlike the original non-equilibrium active systems, the AH systems can be investigated based on equilibrium statistical mechanics. This might lead to a better understanding of phase behavior in active systems such as the motility-induced phase separation (MIPS) \cite{Paoluzzi2022}, at least if one can find a similar behavior in AH systems. Another question is how long-wavelength phonon modes affect collective particle patterns and the stability of active systems. In this context, the stability of solids, be they amorphous or crystalline, is an interesting issue, as has been recently investigated in a simulation of particles performing run-and-tumble motion in a solid. Here, it has been demonstrated that coupling the particles' activity to soft phonon modes can destabilize the solid  \cite{dey2024enhancedlongwavelengthmerminwagner}. Also, in this case, an approach based on AH dynamics may elucidate the activity-phonon coupling based on a statistical mechanics approach that allows the determination of the free energy of the system.

As shown below, AH models lead to non-standard microscopic expressions for basic thermodynamic quantities. For example, in addition to the standard virial terms, the expression of pressure contains terms due to the coupling of spins and velocities. As demonstrated below, these terms can be straightforwardly derived in the framework of Statistical Mechanics. Another example is temperature. While it is generally unrelated to the average kinetic energy for AH systems, it follows directly from the equipartition theorem (see below). Concerning the definition of basic thermodynamic quantities, the situation is very different for active systems where one has to employ empirical concepts such as ``effective temperature'' and introduce pressure via phenomenological definitions \cite{Solon2015}. Here, the study of AH models might help better understand the meaning of basic thermodynamic properties in active systems.

The equations of motion that are obtained from AH models in general and from the Hamilton function of the CTCD model in particular are rather complex, and it is not straightforward to derive a symplectic algorithm with which one can solve the equations numerically in the framework of a molecular dynamics (MD) simulation. In fact, Casiulis {\it et al.}~\cite{Casiulis2020_1, Casiulis2020_2} have used a fourth-order Runge-Kutta scheme to solve the equations of motion since a symplectic integration scheme was not available. However, such a non-symplectic integrator destroys many of the properties of Hamiltonian dynamics, namely the time reversibility, the conservation of energy (one expects a drift of energy with time), and the conservation of phase space volume \cite{Leimkuhler2004}. This can lead to uncontrolled errors and spurious dynamical features.

Another issue is the thermostatting of AH models. These models have uncommon thermodynamic properties, although -- unlike the claims in Ref.~\cite{Casiulis2020_2} -- the equipartition theorem holds for these models. For example, with respect to the temperature control of all the degrees of freedom of the system via a thermostat, as a consequence of the coupling of spins and velocities, the temperature cannot be expressed as an average of the kinetic energy. Thus, ``classical'' thermostats for use in an MD simulation, such as the Nos\'e thermostat \cite{Nose1984_1, Nose1984_2}, have to be rederived for the AH models.

In this work, we propose a state-of-the-art symplectic algorithm for AH models based on the Liouville operator splitting technique. This includes deriving an algorithm for performing MD simulations in the canonical ensemble, i.e.~at constant temperature $T$. To this end, we employ Nos\'e's approach \cite{Nose1984_1, Nose1984_2} based on the extended Lagrangian formalism first introduced by Andersen \cite{Andersen1980}. Nos\'e's thermostat has the problem \cite{Hoover1985} that it leads to equations of motion in a virtual time that is related to the real-time by a dynamic scaling variable. By incorporating a Poincar\'e  time transformation, we solve this problem, as proposed by Bond {\it et al.}~\cite{Bond1999}. As a result, we obtain a symplectic integration scheme to evolve the Nos\'e–Poincar\'e equations of motion within a microcanonical ensemble framework of an extended system that ensures the canonical sampling for the physical system to be thermostatted.

Using our symplectic algorithm, we investigate the phase behavior of the CTCD model where spins $\vec{S}_i$ and velocities $\vec{v}_i$ of each particle $i$ are coupled via terms of the form $K \vec{S}_i \cdot \vec{v}_i$ (with $K$ a coupling constant). Here, we focus on systems at low densities that undergo a finite-size transformation from a fluid phase at high temperature to a ``dynamical cluster phase'' at low temperature, characterized by single clusters moving with a finite center-of-mass velocity. The coupling between velocities and spins tends to destabilize a coherent magnetization of these clusters such that for sufficiently large $N$-particle systems and/or large coupling constants $K$ the average magnetization is strongly reduced in favor of topological defects, as predicted recently \cite{Casiulis2020_2, Bore2016}, such that in the limit $KN^{1/2} \to \infty$ the magnetization vanishes. We note that Cavagna {\it et al.}~\cite{Cavagna2019comment, Casiulis2019reply} have pointed out that the cluster motion, as observed in the CTCD model, is very different from that of real flocks of birds (see also Refs.~\cite{Cavagna2015, Cavagna2018}). So whenever we refer below to the cluster motion in the CTCD model as flocking, we just mean the collective motion of clusters with a non-zero center-of-mass velocity and we do not consider this as a model for flocks of birds.

The rest of the manuscript is organized as follows. First, we introduce the AH model in Sec.~\ref{sec_ahmodel}. Here, we start from a Lagrange function, obtain the Hamilton function from a Legendre transformation, and then write down and discuss Hamilton's equations of motion obtained from the Hamilton function. In Sec.~\ref{sec_thermostat}, we derive Nos\'e-Poincar\'e equations of motion. As a ``by-product'' of this derivation, a formula for thermal energy is obtained that is consistent with the equipartition theorem. This formula and other thermodynamic properties, namely the pressure and the specific heat, are then discussed in Sec.~\ref{sec_thermprop}. Section \ref{sec_algorithm} is devoted to the derivation of the full symplectic algorithm for the MD simulation of (two-dimensional) AH models in the canonical ensemble. In Sec.~\ref{sec_simulation}, we present the results for the MD simulations of the CTCD model. Finally, we summarize and draw conclusions in Sec.~\ref{sec_conclusion}. 

\section{The AH Model \label{sec_ahmodel}}
In this section, we present the general Hamilton function and equations of motion for AH systems of $N$ particles in two dimensions (2D). The basis of our model is an $XY$ spin fluid. Particles interact with each other via a potential $u(r)$, with $r$ the distance between a pair of particles (of course, this potential could be generalized to any multi-body potential). In addition, each particle $i$ carries a spin, given by the unit vector $\vec{S}_i = (\cos(\theta_i), \sin(\theta_i))$ with the angle $-\pi < \theta_i \le \pi$. Neighbouring spin pairs with indices $i$ and $j$ may interact with each other, as described by interaction terms $\propto \vec{S}_i \cdot \vec{S}_j$. The ``activity'' in the $XY$ spin fluid is introduced through the coupling between spins $\{\vec{S}_i\}$ and velocities $\{\dot{\vec{r}}_i\}$ of the particles via a vector potential $\vec{A}_i$ for each particle $i$, depending in general on the spins $\{\vec{S}_i\}$ and the particle positions $\{\vec{r}_i\}$ ($i = 1, \dots, N$). The form of the vector potential remains very general in this context, and a suitable choice of it will also reproduce charged particles in a magnetic field. For example, by making the spin-spin interaction to zero and choosing a correct vector potential that represents a constant magnetic field, we obtain a system of particles in a magnetic field, interacting via a pair potential $u(r)$.

{\bf Lagrange function.} 
With the above assumptions about the interaction potential, spin-spin interactions and spin-velocity coupling, we obtain the Lagrange function $\mathcal{L}$, given by
\begin{eqnarray}
\mathcal{L} & = & \sum_{i=1}^N \frac{m}{2} \dot{\vec{r}}_i^{\, 2} 
+ \sum_{i=1}^N \frac{I}{2} \dot{\theta}_i^2 
+ \sum_{i=1}^{N} \vec{A}_i \cdot \dot{\vec{r}}_i \nonumber \\ 
& & - \frac{1}{2} \sum_{\substack{i,j \\ i \ne j}} u(r_{ij}) 
+ \frac{1}{2} J \sum_{\substack{i,j \\ i \ne j}} g(r_{ij}) \, \vec{S}_i \cdot \vec{S}_j \, ,
\label{eq_lagrangian}
\end{eqnarray}

with $r_{ij} = |\vec{r}_i - \vec{r}_j|$ the distance between particles $i$ and $j$, $\dot{\theta}_i$ the angular velocity of particle $i$ associated with its spin, $m$ the mass and $I$ the moment of inertia of a particle. The amplitude of the coupling between spin pairs is given by $J g(r_{ij})$ where $J$ is a constant with the unit of energy and the function $g(r_{ij})$, to be specified below, controls the interaction range of spin pairs. 

{\bf Hamilton function.} The canonical momenta associated with the Lagrange function, Eq.~(\ref{eq_lagrangian}), are
\begin{equation}
\begin{split}
\vec{p}_i &= \frac{\partial \mathcal{L}}{\partial \dot{\vec{r}}_i} = m \dot{\vec{r}}_i + \vec{A}_i \\
\omega_i &= \frac{\partial \mathcal{L}}{\partial \dot{\theta}_i} = I \dot{\theta}_i \, ,
\end{split}
\end{equation}
with $i = 1, \dots, N$. The Hamilton function $\mathcal{H}$ is determined from the canonical momenta via Legendre transformation,
\begin{equation}
\mathcal{H} =
\sum_{i=1}^N \left( \vec{p}_i \cdot \dot{\vec{r}}_i + \omega_i \dot{\theta}_i\right) 
-\mathcal{L} \, ,
\end{equation}
from which one obtains 
\begin{eqnarray}
\mathcal{H} & = & 
\sum_{i=1}^N \frac{(\vec{p}_i-\vec{A}_i)^2}{2m} + \sum_{i=1}^N \frac{{\omega_i}^2}{2 I} \nonumber \\
& & + \frac{1}{2} \sum_{\substack{ i,j \\ {i\ne j}}} u(r_{ij})
-\frac{1}{2}J \sum_{\substack{i,j \\ {i\ne j}}} g(r_{ij}) \, \cos(\theta_{ij}) \, ,
\label{eq_hamiltonian}
\end{eqnarray}
where $\cos(\theta_{ij}) \equiv \vec{S}_i \cdot \vec{S}_j$ with $\theta_{ij} = \theta_i - \theta_j$.

{\bf Equations of motion.} From the Hamilton function (\ref{eq_hamiltonian}), we can now derive the equations of motion for the variables $\omega_i$, $\theta_i$, $\vec{r}_i = (x_i, y_i)$ and $\vec{p}_i = (p_i^x, p_i^y)$ of each particle $i$. The velocity of particle $i$, $\dot{\vec{r}}_i$ is given by 
\begin{equation}
\dot{\vec{r}}_i = \frac{\partial \mathcal{H}}{\partial \vec{p}_i} = 
\frac{\vec{p}_i-\vec{A}_i}{m} \, .
\label{eq_eom_dr}
\end{equation}
The scalar variables $\omega_i$ and $\theta_i$ follow the equations of motion $\dot{\theta}_i = \frac{\partial \mathcal{H}}{\partial \omega_i}$ and $\dot{\omega}_i = - \frac{\partial \mathcal{H}}{\partial \theta_i}$, respectively, and thus we obtain
\begin{eqnarray}
\dot{\theta}_i & = & \frac{\omega_i}{I} \label{eq_eom_dtheta}\\
\dot{\omega}_i & = & \sum_j(\vec{p}_j-\vec{A}_j)\cdot\frac{\partial \vec{A}_j}{\partial \theta_i}-J\sum_{\substack{j\\j\ne i}}g(r_{ij})\sin\theta_{ij} \, . \label{eq_eom_domega}
\end{eqnarray}
Note that in Eq.~(\ref{eq_eom_domega}) the terms $(\vec{p}_j-\vec{A}_j)\cdot\frac{\partial \vec{A}_j}{\partial \theta_i}=m\dot{\vec{r}}_j \cdot\frac{\partial \vec{A}_j}{\partial \theta_i}$ are due to the coupling of spins and velocities. The equations $\dot{p}_i^x = - \frac{\partial \mathcal{H}}{\partial x_i}$ and $\dot{p}_i^y = - \frac{\partial \mathcal{H}}{\partial y_i}$ yield
\begin{eqnarray}
\dot{p}_i^x & = & \sum_{\substack{j}}\left[(p_j^x-A_j^x) \frac{\partial A_j^x}{\partial x_i} 
 + ({p}_j^y-{A}_j^y)\frac{\partial A_j^y}{\partial x_i} \right] \nonumber \\
 & & -\sum_{\substack{j\\j\ne i}}\frac{\partial u(r_{ij})}{\partial x_i}
 +J\sum_{\substack{j\\j\ne i}}\frac{\partial g(r_{ij})}{\partial x_i}\cos\theta_{ij} 
 \label{eq_eom_dpx} \\
\dot{p}_i^y & = & \sum_{\substack{j}}\left[(p_j^x-A_j^x) \frac{\partial A_j^x}{\partial y_i} 
+ ({p}_j^y-{A}_j^y)\frac{\partial A_j^y}{\partial y_i} \right] \nonumber \\
& & -\sum_{\substack{j\\j\ne i}}\frac{\partial u(r_{ij})}{\partial y_i}
+J\sum_{\substack{j\\j\ne i}}\frac{\partial g(r_{ij})}{\partial y_i}\cos\theta_{ij} \, .
\label{eq_eom_dpy}
\end{eqnarray}
The terms $m\dot{r}_k^\alpha \frac{\partial A_k^\alpha}{\partial \eta_l}$ with $\alpha = x, y$ and $\eta_l = x_l, y_l$ in Eqs.~(\ref{eq_eom_dpx}) and (\ref{eq_eom_dpy}) are non-zero if the vector potential depends on the spatial coordinates. These terms as well as the terms $m\dot{\vec{r}}_j \cdot\frac{\partial \vec{A}_j}{\partial \theta_i}$ in Eq.~(\ref{eq_eom_dtheta}) require special attention when one aims at coupling the Hamiltonian system, as defined by the Hamilton function (\ref{eq_hamiltonian}), to a thermostat. In the next section, we derive the equations of motion for a thermostatted version of our AH model that provides a correct sampling in the canonical ensemble. To this end, we follow the approach of Nosé, augmented by a Poincaré transformation to obtain equations of motion in real time (see below). We therefore refer to this thermostat as the Nosé-Poincaré thermostat.

\section{The Nosé-Poincaré thermostat \label{sec_thermostat}}
Nos\'e's approach to thermostat a system in the framework of a molecular dynamics (MD) simulation \cite{Nose1984_1, Nose1984_2} is based on the extended Lagrangian method, first proposed by Andersen \cite{Andersen1980} in the context of MD simulations at constant pressure. In Nos\'e's thermostat, the temperature of an $N$ particle system is controlled via a scaling variable $s$ that couples to the velocities of the particles without affecting their positions. Since the velocities are given by time derivatives of the positions, one has to couple the variable $s$ to the time differential $\mathrm{d}t$ such that one obtains a ``virtual'' time differential $\mathrm{d}t^{\prime}= s \mathrm{d}t$. The relation of the velocities of particle $i$ with respect to virtual and real time is thus given by $\dot{\vec{r}}^{\, \prime}_i = \frac{\mathrm{d} \vec{r_i}}{\mathrm{d} t^\prime} = \frac{1}{s}\frac{\mathrm{d} \vec{r}_i}{\mathrm{d} t}$ and $\dot{\theta}^{\, \prime}_i = \frac{\mathrm{d} \theta^{\, \prime}_i}{\mathrm{d} t^{\, \prime}} = \frac{1}{s}\frac{\mathrm{d} \theta_i}{\mathrm{d} t}$. The variable $s$ is a dimensionless dynamical variable that is associated with a kinetic and a potential energy. It assumes the role of a heat bath that keeps the temperature of the physical $N$-particle system constant via a dynamical scaling of the velocities. While the physical system plus the additional degrees of freedom due to $s$ may define an isolated system in the microcanonical ensemble, after ``integrating out'' the variable $s$ and its momentum from thermodynamic averages one obtains the physical system being in a canonical ensemble (see below).

The starting point of Nos\'e's approach is to set up an extended Lagrange function that introduces the dynamical variable $s$ and its coupling to the velocities of the particles. In our case it has the following form:
\begin{widetext}
	\begin{equation}
	\mathcal{L}_{\text{Nos\'e}} =  \sum_{i=1}^N \frac{m}{2} s^2 (\dot{\vec{r}}^{\, \prime}_i)^2 
	+ \sum_{i=1}^N \frac{I}{2} s^2 (\dot{\theta}_i^{\, \prime})^2 
	+ \sum_{i=1}^N \vec{A}_i \cdot s \dot{\vec{r}}^{\, \prime}_i 
	- \frac{1}{2} \sum_{\substack{i,j \\ i \ne j}} u(r_{ij}) 
	+ \frac{1}{2} J \sum_{\substack{i,j \\ i \ne j}} g(r_{ij}) \, \vec{S}_i \cdot \vec{S}_j 
	+ \frac{1}{2} Q \dot{s}^2 - \frac{C}{\beta} \ln(s) \, ,
	\label{nose_lagrangian}
	\end{equation}
\end{widetext}

Here, the variable $s$ is associated with a kinetic energy $\frac{1}{2} Q \dot{s}^2$ (with $Q$ an effective ``mass'') and a potential energy $\frac{C}{\beta} \ln(s)$ (with $C$ a constant and $\beta = 1/(k_{\mathrm{B}}T)$ the inverse thermal energy). As we shall see below, the logarithmic form of the potential provides a correct sampling in the canonical ensemble with respect to the physical system.

By performing a Legendre transformation of the extended Lagrangian (\ref{nose_lagrangian}), the Nos\'e-Hamiltonian follows as
\begin{align}
	\mathcal{H}_{\text{Nos\'e}} &= \sum_{i=1}^N \frac{(\vec{p}^{\, \prime}_i - s\vec{A}_i)^2}{2ms^2} 
	+ \sum_{i=1}^N \frac{(\omega^{\, \prime}_i)^2}{2Is^2} 
	+ \frac{1}{2} \sum_{\substack{i,j \\ i\ne j}} u(r_{ij}) \nonumber \\
	&\quad - \frac{1}{2}J \sum_{\substack{i,j \\ i\ne j}} g(r_{ij}) \cos(\theta_{ij})
	+ \frac{p_s^2}{2Q} + \frac{C}{\beta}\ln(s)
	\label{nose_hamiltonian}
	\end{align}
where, $\vec{p}^{\, \prime}_i = s\vec{p}_i$ and $\omega^{\, \prime}_i = s\omega_i$. An unpleasant feature of the Nos\'e-Hamiltonian (\ref{nose_hamiltonian}) is associated with the fact that the resulting equations of motion are given with respect to virtual time. Hoover \cite{Hoover1985} addressed this issue by reformulating the equations of motion in real time. However, in his approach, the equations of motion are rewritten via a non-canonical change of variables. While Hoover's equations produce samples that are canonically distributed and evolve in real time, they cannot be derived from any Hamilton function and thus do not possess a symplectic structure.

To obtain a Hamilton function that leads to equations of motion in real time, a Poincaré transformation \cite{zare1975time, Bond1999} of the Hamiltonian can be applied. To illustrate the meaning of this transformation, consider a Hamiltonian for a system with four degrees of freedom $\mathcal{H}=\mathcal{H}(q, p^{\, \prime}, s, p_s)$. Here, $p^{\, \prime}= s p$ is the relation between the momentum $p^{\, \prime}$ in virtual time $t^{\, \prime}$ and the momentum $p$ in real time $t$. The scaling variable $s>0$ is given by $s=\frac{\mathrm{d} t^\prime}{\mathrm{d} t}$. Now, we define the Hamiltonian  
\begin{equation}
    \widetilde{\mathcal{H}}=s \, (\mathcal{H}(q, p^{\, \prime}, s, p_s) - \mathcal{H}_0),
\end{equation}
where $\mathcal{H}_0$ corresponds to the initial value of $\mathcal{H}$. The equation of motion with respect to time $t$ that are obtained from $\widetilde{\mathcal{H}}$ are given by
\begin{eqnarray}
    \frac{\mathrm{d} q}{\mathrm{d} t}&=&s\frac{\partial \mathcal{H}}{\partial p^{\, \prime}} + (\mathcal{H}-\mathcal{H}_0)\frac{\partial s}{\partial p^{\, \prime}} \label{eq_pt1}\\
    \frac{\mathrm{d} p^{\, \prime}}{\mathrm{d} t}&=&-s\frac{\partial \mathcal{H}}{\partial q} - (\mathcal{H}-\mathcal{H}_0)\frac{\partial s}{\partial q} \label{eq_pt2}\\
    \frac{\mathrm{d} s}{\mathrm{d} t}&=&s\frac{\partial \mathcal{H}}{\partial p_s} + (\mathcal{H}-\mathcal{H}_0)\frac{\partial s}{\partial p_s} \label{eq_pt3}\\
    \frac{\mathrm{d} p_s}{\mathrm{d} t}&=&-s\frac{\partial \mathcal{H}}{\partial s} - (\mathcal{H}-\mathcal{H}_0) \label{eq_pt4} \, .
\end{eqnarray}

As is evident from Eqs.~(\ref{eq_pt1})-(\ref{eq_pt4}), along the constant energy surface, $\mathcal{H}=\mathcal{H}_0$, the dynamics of the transformed system in real time is equivalent to that of the original system that evolves in virtual time. Note, however, that due to discretization errors for any discretized version of Eqs.~(\ref{eq_pt1})-(\ref{eq_pt4}) the equation $\mathcal{H}=\mathcal{H}_0$ no longer holds and one has to take into account the additional terms $\propto (\mathcal{H}-\mathcal{H}_0)$.

If we apply the Poincaré transformation to the Nos\'e-Hamiltonian, we obtain
\begin{eqnarray}
\widetilde{\mathcal{H}} &=& s\left(\mathcal{H}_{\text{Nos\'e}} - \mathcal{H}_0\right) \nonumber\\
&=& s \biggl( \sum_{i=1}^N \frac{(\vec{p}^{\, \prime}_i - s\vec{A}_i)^2}{2ms^2} + \sum_{i=1}^N \frac{(\omega^{\, \prime}_i)^2}{2Is^2} + \frac{1}{2} \sum_{\substack{ i,j \\ i \ne j}} u(r_{ij}) \nonumber \\
& & - \frac{1}{2} J \sum_{\substack{i,j \\ i \ne j}} g(r_{ij}) \cos(\theta_{ij}) + \frac{p_s^2}{2Q} + \frac{C}{\beta} \ln(s) - \mathcal{H}_0 \biggr) \nonumber\\
\label{np_hamiltonian}
\end{eqnarray}

In the following, we use the notation $\mathcal{R} = (\vec{r}_1, \dots, \vec{r}_N)^T$, $\mathcal{P} = (\vec{p}_1, \dots, \vec{p}_N)^T$, $\Theta = (\theta_1, \dots, \theta_N)^T$, and  $\Omega = (\omega_1, \dots, \omega_N)^T$. We shall see that, for the variables $\left(\mathcal{R}, \mathcal{P}, \Theta, \Omega \right)$ of the physical system, the Nos\'e-Poincar\'e-Hamiltonian, $\widetilde{\mathcal{H}}$, provides a sampling in the canonical ensemble.
 
Consider the probability of finding a particular configuration in the phase space described by the variables $\left(\mathcal{R}, \mathcal{P}, \Theta, \Omega \right)$ of the physical system:
\begin{eqnarray}    
    &&\mathrm{d}\mathcal{R}\mathrm{d}\mathcal{P}\mathrm{d}\Theta \mathrm{d}\Omega \, \mathcal{F}\left(\mathcal{R}, \mathcal{P}, \Theta, \Omega \right) \equiv \int \mathrm{d}p_s \int \mathrm{d}s \times \nonumber \\
    && \times
    \mathrm{d}\mathcal{R}\mathrm{d}\mathcal{P}^{\, \prime}\mathrm{d}\Theta \mathrm{d}\Omega^{\, \prime} \, \mathcal{F}_{\rm ext} \left(\mathcal{R}, \mathcal{P}^{\, \prime}, \Theta, \Omega^{\, \prime}, s, p_s \right) \label{eq_probmist}
\end{eqnarray}
with $\mathcal{F}$ and $\mathcal{F}_{\rm ext}$ the probability density of the physical and the extended system, respectively. For the Nos\'e-Poincar\'e-Hamiltonian, $\widetilde{\mathcal{H}}$, we can write the probability of finding a particular configuration of energy $\widetilde{\mathcal{H}}_0$ within the microcanonical ensemble of the extended phase space $\left( \mathcal{R}, \mathcal{P}^{\, \prime}, \Theta, \Omega^{\, \prime}, s, p_s \right)$ as
\begin{eqnarray}
    &&\mathrm{d}p_s \mathrm{d}s \mathrm{d}\mathcal{P}^{\, \prime} \mathrm{d}\mathcal{R} \mathrm{d}\Omega^{\, \prime} \mathrm{d}\Theta \, \mathcal{F}_{\rm ext} \left(\mathcal{R}, \mathcal{P}^{\, \prime}, \Theta, \Omega^{\, \prime}, s, p_s \right) = \nonumber \\
    &&\frac{\mathrm{d}p_s \mathrm{d}s \mathrm{d}\mathcal{P}^{\, \prime} \mathrm{d}\mathcal{R} \mathrm{d}\Omega^{\, \prime} \mathrm{d}\Theta \, \delta \left( \widetilde{\mathcal{H}}-\widetilde{\mathcal{H}}_0\right)}{\int \mathrm{d}p_s \int \mathrm{d}s \int \mathrm{d}\mathcal{P}^{\, \prime} \int \mathrm{d}\mathcal{R} \int \mathrm{d}\Omega^{\, \prime} \int \mathrm{d}\Theta \, \delta \left( \widetilde{\mathcal{H}}-\widetilde{\mathcal{H}}_0\right)} \, . \label{eq_probmist2}
\end{eqnarray}
Inserting Eq.~(\ref{eq_probmist2}) in (\ref{eq_probmist}), we obtain
\begin{eqnarray}
    && \mathrm{d}\mathcal{P} \mathrm{d}\mathcal{R} \mathrm{d}\Omega \mathrm{d}\Theta \, \mathcal{F} \left(\mathcal{R}, \mathcal{P}, \Theta, \Omega \right) = \nonumber \\
    &&\frac{1}{\widetilde{\mathcal{Z}} N! h^{N_f}}{\int \mathrm{d}p_s \int \mathrm{d}s  \mathrm{d}\mathcal{P}^{\, \prime} \mathrm{d}\mathcal{R} \mathrm{d}\Omega^{\, \prime} \mathrm{d}\Theta \, \delta \left( \widetilde{\mathcal{H}}-\widetilde{\mathcal{H}}_0\right)} \, ,
\end{eqnarray}
where $h$ is Planck's constant, $N_f$ the number of degrees of freedom, and $\widetilde{\mathcal{Z}}$ the partition function, given by
\begin{eqnarray}
    \widetilde{\mathcal{Z}} &=& \frac{1}{N! h^{N_f}}\int \mathrm{d}p_s \int \mathrm{d}s \int \mathrm{d}\mathcal{P}^{\, \prime} \int \mathrm{d}\mathcal{R} \int \mathrm{d}\Omega^{\, \prime} \int \mathrm{d}\Theta \times \nonumber \\
    && \times \delta \left( \widetilde{\mathcal{H}}-\widetilde{\mathcal{H}}_0\right)
\end{eqnarray}
Using $\widetilde{\mathcal{H}}_0=0$ and expanding $\widetilde{\mathcal{H}}$, we obtain
\begin{eqnarray}
&& \mathrm{d}\mathcal{P} \mathrm{d}\mathcal{R} \mathrm{d}\Omega \mathrm{d}\Theta \, \mathcal{F} \left(\mathcal{R}, \mathcal{P}, \Theta, \Omega \right) = \\
&&\frac{1}{\widetilde{\mathcal{Z}} N! h^{N_f}} \int \mathrm{d}p_s \int \mathrm{d}s  \, \mathrm{d}\mathcal{P}^{\, \prime} \mathrm{d}\mathcal{R} \mathrm{d}\Omega^{\, \prime} \mathrm{d}\Theta \, \delta \left[ s\left(\mathcal{H}_{\text{Nos\'e}} - \mathcal{H}_0\right)\right] \nonumber
\end{eqnarray}
Since $s$ is strictly positive, we can make the change of variables $\vec{p}^{\, \prime}_i \rightarrow s \vec{p_i}$ and $\omega^{\, \prime}_i \rightarrow s \omega_i$, yielding
\begin{eqnarray}
 && \mathrm{d}\mathcal{P} \mathrm{d}\mathcal{R} \mathrm{d}\Omega \mathrm{d}\Theta \, \mathcal{F} \left(\mathcal{R}, \mathcal{P}, \Theta, \Omega \right) = \nonumber \\
 &&\frac{1}{\widetilde{\mathcal{Z}}N! h^{N_f}} \int \mathrm{d}p_s \int \mathrm{d}s \, \mathrm{d}\mathcal{P} \mathrm{d}\mathcal{R} \mathrm{d}\Omega \mathrm{d}\Theta \; s^{N_f} \times \\
    && \times \delta \left[ s\left(\mathcal{H}(\mathcal{P}, \mathcal{R},\Omega, \Theta)+\frac{p_s^2}{2Q}+\frac{C}{\beta}\ln(s)-\mathcal{H}_0\right)\right] \nonumber
\end{eqnarray}
Now by using the identity, $\delta \left[ r(s)\right] = \delta (s-s_0)/r^{\prime}(s_0)$, with $s_0$ the simple root of $r(s)$ and $r^{\prime}(s_0)$ the derivative of $r(s)$ with respect to $s$ at $s_0$, we obtain
\begin{widetext}
\begin{eqnarray}
    \mathrm{d}\mathcal{R} \mathrm{d}\mathcal{P} \mathrm{d}\Omega \mathrm{d}\Theta \mathcal{F} &=&\frac{1}{\widetilde{\mathcal{Z}}N! h^{N_f}}\int \mathrm{d}p_s \int \mathrm{d}s \mathrm{d}\mathcal{R} \mathrm{d}\mathcal{P} \mathrm{d}\Omega \mathrm{d}\Theta \frac{s^{N_f}}{Ck_B T}\delta\left[ 
s- \exp\left( -\frac{1}{Ck_B T}\left(\mathcal{H}(\mathcal{R},\mathcal{P}, \Omega, \Theta) +\frac{p_s^2}{2Q}-\mathcal{H}_0 \right) \right)\right] \nonumber \\
&=& \frac{1}{\widetilde{\mathcal{Z}}N! h^{N_f}C k_B T} \; \exp\left(\frac{N_f \mathcal{H}_0}{C k_B T}\right) \; \int \mathrm{d}p_s \mathrm{d}\mathcal{R} \mathrm{d}\mathcal{P} \mathrm{d}\Omega \mathrm{d}\Theta \; \exp\left( -\frac{N_f}{C k_B T} \left( \mathcal{H} +\frac{p_s^2}{2Q} - \mathcal{H}_0 \right) \right) \nonumber \\
&=& \frac{\mathcal{N}}{\widetilde{\mathcal{Z}}N! h^{N_f}} \mathrm{d}\mathcal{R} \mathrm{d}\mathcal{P} \mathrm{d}\Omega \mathrm{d}\Theta \exp \left( -\frac{N_f}{C k_B T} \mathcal{H}\left(\mathcal{R}, \mathcal{P}, \Theta, \Omega \right)\right)
\end{eqnarray}
\end{widetext}
with $\mathcal{N}$ being a constant. By choosing $C=N_f$, the probability of finding a configuration of the physical system is
\begin{eqnarray}
    &&\mathrm{d}\mathcal{R} \mathrm{d}\mathcal{P} \mathrm{d}\Omega \mathrm{d}\Theta \; \mathcal{F} \nonumber \\ &&=\frac{\mathrm{d}\mathcal{R} \mathrm{d}\mathcal{P} \mathrm{d}\Omega \mathrm{d}\Theta \; \exp\left( -\frac{\mathcal{H}}{k_B T} \right)}{\int \mathrm{d}\mathcal{R} \int \mathrm{d}\mathcal{P} \int \mathrm{d}\Omega \int \mathrm{d}\Theta \; \exp\left( -\frac{\mathcal{H}}{k_B T} \right)} \, ,
\end{eqnarray}
and thus the sampling of configurations of the physical system is according to the canonical ensemble.

\textbf{Equations of motion:} From the Hamilton function (\ref{np_hamiltonian}), we can now derive the equations of motion for the variables $\omega_i$, $\theta_i$, $\vec{r}_i = (x_i, y_i)$ and $\vec{p}^{\, \prime}_i = (p_i^{\, \prime x}, p_i^{\, \prime y})$ of each particle $i$. The velocity of particle $i$ is given by
\begin{equation}
\dot{\vec{r}}_i = \frac{\partial \widetilde{\mathcal{H}}}{\partial \vec{p}_i^{\, \prime}} = \frac{\vec{p}_i^{\, \prime} -s\vec{A}_i}{ms}
\label{np_eom_dr}
\end{equation}
The equations of motion of the scalar variables $\theta_i$ and $\omega_i^{\, \prime}$ follow respectively from $\dot{\theta}_i = \frac{\partial \widetilde{\mathcal{H}}}{\partial \omega_i^{\, \prime}}$ and $\dot{\omega}_i^{\, \prime} = -\frac{\partial \widetilde{\mathcal{H}}}{\partial \theta_i}$, and thus we obtain
\begin{eqnarray}
\dot{\theta}_i & = & \frac{\omega_i^{\, \prime}}{Is} \label{np_eom_dtheta}\\
\dot{\omega}_i^{\, \prime} & = &
s \biggl[ \sum_j\frac{\vec{p}_j^{\, \prime}-s\vec{A}_j}{ms}\cdot\frac{\partial \vec{A}_j}{\partial \theta_i}-J\sum_{\substack{j\\j\ne i}}g(r_{ij})\sin\theta_{ij} \biggr] . \label{np_eom_domega}
\end{eqnarray}
The equations $\dot{p}_i^{\, \prime x} = - \frac{\partial \widetilde{\mathcal{H}}}{\partial x_i}$ and $\dot{p}_i^{\, \prime y} = - \frac{\partial \widetilde{\mathcal{H}}}{\partial y_i}$ yield
\begin{eqnarray}
\dot{p}_i^{\, \prime x} & = & s\biggl[\sum_{\substack{j}}\left(\frac{p_j^{\, \prime x}-sA_j^x}{ms} \frac{\partial A_j^x}{\partial x_i} 
 + \frac{{p}_j^{\, \prime y}-s{A}_j^y}{ms}\frac{\partial A_j^y}{\partial x_i} \right) \nonumber \\
 & & -\sum_{\substack{j\\j\ne i}}\frac{\partial u(r_{ij})}{\partial x_i}
 +J\sum_{\substack{j\\j\ne i}}\frac{\partial g(r_{ij})}{\partial x_i}\cos\theta_{ij} \biggr] 
 \label{np_eom_dpx}
 \end{eqnarray}
 
 \begin{eqnarray}
\dot{p}_i^{\, \prime y} & = & s\biggl[\sum_{\substack{j}}\left(\frac{p_j^{\, \prime x}-sA_j^x}{ms} \frac{\partial A_j^x}{\partial y_i} 
 + \frac{{p}_j^{\, \prime y}-s{A}_j^y}{ms}\frac{\partial A_j^y}{\partial y_i} \right) \nonumber \\
 & & -\sum_{\substack{j\\j\ne i}}\frac{\partial u(r_{ij})}{\partial y_i}
 +J\sum_{\substack{j\\j\ne i}}\frac{\partial g(r_{ij})}{\partial y_i}\cos\theta_{ij} \biggr] .
\label{np_eom_dpy}
\end{eqnarray}
The time evolution of the scaling variable $s$ is
\begin{equation}
    \dot{s} = \frac{\partial \widetilde{\mathcal{H}}}{\partial p_s} = s \frac{p_s}{Q}
    \label{np_eom_s}
\end{equation}
and that of the corresponding momentum, $\dot{p}_s = -\frac{\partial \widetilde{\mathcal{H}}}{\partial s}$ is given by 

\begin{eqnarray}
\dot{p}_s &=& \sum_{i=1}^N \left( \frac{\vec{p}_i^{\, \prime 2}}{ms^2} - \frac{\vec{p}_i^{\, \prime} \cdot \vec{A}_i}{ms} + \frac{\omega_i^{\, \prime 2}}{Is^2} \right) \nonumber \\
&& - 3Nk_BT - \Delta \widetilde{\mathcal{H}}\left( \mathcal{R}, \mathcal{P}, \Theta, \Omega, s, p_s \right)
\label{np_eom_dps}
\end{eqnarray}
where
\begin{eqnarray}
\Delta \widetilde{\mathcal{H}} &=& \sum_{i=1}^N \frac{(\vec{p}_i^{\, \prime} - s\vec{A}_i)^2}{2ms^2} + \sum_{i=1}^N \frac{\omega_i^{\, \prime 2}}{2Is^2} + \frac{1}{2} \sum_{\substack{i,j \\ i \ne j}} u(r_{ij}) \\
&-& \frac{J}{2} \sum_{\substack{i,j \\ i \ne j}} g(r_{ij}) \cos(\theta_{ij}) + \frac{p_s^2}{2Q} + 3Nk_B T \ln(s) - \mathcal{H}_0 \nonumber
\label{delta_H}
\end{eqnarray}
The value of $\mathcal{H}_0$ is chosen such that $\Delta \widetilde{\mathcal{H}}\left(\mathcal{R}_0, \mathcal{P}_0, \Theta_0, \Omega_0, s_0, p_{s_0} \right) = 0$.

Equation (\ref{np_eom_dps}) describes a restoring force with respect to the ``instantaneous'' temperature of the system. It implies a non-standard microscopic definition of temperature to be discussed in the next section where we also derive the ensemble averages of other thermodynamic properties of AH systems, namely the pressure and the specific heat. 
\section{Thermodynamic properties \label{sec_thermprop}}
In this section, we discuss the thermodynamics of systems that are described by the Hamilton function $\mathcal{H}$, as defined by Eq.~(\ref{eq_hamiltonian}). We show how one can express temperature $T$, pressure $P$, and specific heat $C$ as canonical averages of an observable. For our system with Hamilton function $\mathcal{H}$ the canonical average of an observable $A$ is defined by
\begin{equation}
 \langle A \rangle = \frac{\mathcal{N}}{\mathcal{Z}} \int \mathrm{d} \mathcal{P} \int \mathrm{d}\mathcal{R} \int \mathrm{d} \Theta \int \mathrm{d} \Omega \; A \, \exp \left(-\beta \mathcal{H} \right)  
\end{equation}
with $\mathcal{N}$ a normalization constant that is irrelevant for the following. The partition function $\mathcal{Z}$ is 
\begin{equation}
  \mathcal{Z} = \mathcal{N} \int \mathrm{d} \mathcal{P} \int \mathrm{d}\mathcal{R} \int \mathrm{d} \Theta \int \mathrm{d} \Omega \; \exp \left(-\beta \mathcal{H} \right) \, .
\end{equation}
We shall see below that due to the coupling of velocities to vector potentials, non-standard formulae for thermodynamic properties are obtained.

{\bf Temperature.} The kinetic energy, $\mathcal{T}$, that enters the Hamilton function (\ref{eq_hamiltonian}) is given by $\mathcal{T} = \sum_i (\vec{p}_i - \vec{A}_i)^2/(2m) + \sum_i \omega_i^2/(2I)$. For a standard Hamiltonian system, the average kinetic energy is related to the thermal energy by
\begin{equation}
\langle \mathcal{T} \rangle = \frac{N_f}{2} k_B T \, .
\label{eq_equipart}
\end{equation}

In our system, due to the coupling of vector potential and velocity, Eq.~(\ref{eq_equipart}) no longer holds. Of course, the average rotational energy still follows the standard formula, $\langle \sum_i \omega_i^2/I \rangle = N k_B T$. From our analysis of the Nos\'e-Poincar\'e thermostat, we can infer the following relation from Eq.~(\ref{np_eom_dps}):
\begin{equation}
    2 N k_B T = \left\langle \sum_{i=1}^N \left( \frac{\vec{p}_i^{\; 2}}{m} - \frac{\vec{p}_i \cdot \vec{A}_i}{m} \right) \right\rangle \, .
\end{equation}
This formula is -- as it should be -- consistent with the equipartition theorem \cite{Huang2008},
\begin{equation}
    \left\langle x^\alpha \frac{\partial \mathcal{H}}{\partial x^{\beta}} \right\rangle = \delta_{\alpha \beta} k_{\mathrm{B}} T \, ,
\end{equation}
where $x^{\alpha}$ is the $\alpha$'th component of a phase space variable $\vec{x}$ (e.g.~$\vec{x}=\vec{p}_i$). Thus, for our system the temperature $T$ can be written as
\begin{equation}
    T = \frac{1}{3Nk_B} \left\langle \sum_{i=1}^N \left( \frac{\vec{p}_i^{\; 2}}{m} - \frac{\vec{p}_i \cdot \vec{A}_i}{m} + \frac{\omega_i^2}{I} \right) \right\rangle \, .
\label{eq_temperature}
\end{equation}
This formula clearly shows that the temperature $T$ is not proportional to the average kinetic energy. Instead of terms $\frac{1}{m} (\vec{p}_i - \vec{A}_i)^2$ for a particle $i$, the corresponding terms in Eq.~(\ref{eq_temperature}) are given by $\frac{1}{m} \vec{p}_i \cdot (\vec{p}_i - \vec{A}_i)$. Below, we rationalize Eq.~(\ref{eq_temperature}) using MD simulations where the system is coupled to a Nos\'e-Poincar\'e thermostat.

{\bf Pressure.} Also the pressure is affected by the coupling of velocities and vector potentials in a non-standard manner. For a two-dimensional system, the pressure can be defined by the derivative of the Helmholtz free energy $F(N, V_{2d}, T)$ with respect to the area $V_{2d}$,
\begin{equation}
P = - \left. \frac{\partial F}{\partial V_{2d}} \right|_{N, T} = k_B T \frac{\partial \ln \mathcal{Z}}{\partial V_{2d}} = \frac{k_B T}{\mathcal{Z}} \frac{\partial \mathcal{Z}}{\partial V_{2d}}
\end{equation}
Consider a system in a square of area $V_{2d} = L^2$. We define scaled coordinates $\vec{\phi}_i$ as $\vec{\phi}_i = \frac{\vec{r}_i}{L}$ such that
\begin{equation}
\mathrm{d} \mathcal{R} \equiv \mathrm{d} \vec{r}_1 \mathrm{d} \vec{r}_2 \dots \mathrm{d} \vec{r}_N =  V_{2d}^N \mathrm{d} \vec{\phi}_1 \mathrm{d} \vec{\phi}_2 \dots \mathrm{d} \vec{\phi}_N
\end{equation}
Similarly, $u(\{ \vec{r}_i \}) = u(\{ \vec{\phi}_i \}, V_{2d}) $, $g(\{ \vec{r}_i \}) = g(\{ \vec{\phi}_i \}, V_{2d}) $ and $\vec{A}_i(\{ \vec{r}_i \}) = \vec{A}_i(\{ \vec{\phi}_i \}, V_{2d}) $. Now all the phase space coordinates can be collected in a single $6N$-dimensional vector as $\vec{\mu} = (\mathcal{P}, \Phi, \Omega, \Theta)^T$ with $\Phi = (\vec{\phi}_1, \dots, \vec{\phi}_N)$. Then the generalised form of pressure is given by
\begin{eqnarray}
    P &=& k_B T \frac{\int \mathrm{d} \vec{\mu} \left( NV_{2d}^{N-1} - V_{2d}^N \beta \frac{\partial \mathcal{H}}{\partial V_{2d}} \right){\rm e}^{-\beta \mathcal{H}}}{V_{2d}^N \int \mathrm{d} \vec{\mu} \; {\rm e}^{-\beta \mathcal{H}}} \nonumber \\
&=& k_B T \frac{N}{V_{2d}} + P_{\rm act} + P_{\rm pair} \, ,
\label{pressure}
\end{eqnarray}
where $P_{\rm act}$ and $P_{\rm pair}$ are respectively defined by
\begin{equation}
P_{\rm act} = \left< \sum_{i=1}^N (p_i^x - A_i^x) \frac{\partial A_i^x}{\partial V_{2d}} + \sum_{i=1}^N (p_i^y - A_i^y) \frac{\partial A_i^y}{\partial V_{2d}} \right>
\label{eq_press1}
\end{equation}
and
\begin{equation}
P_{\rm pair} = \frac{-1}{4V_{2d}} \left< \sum_{\substack{i,j \\ {i\ne j}}} \left[ \frac{\partial u(r_{ij})}{\partial r_{ij}} 
 - J \frac{\partial g(r_{ij})}{\partial r_{ij}} \cos(\theta_{ij}) \right] r_{ij} \right> \, .
 \label{eq_press2}
\end{equation}
In the framework of Hamiltonian dynamics, we define an active pressure via Eqs.~(\ref{pressure})-(\ref{eq_press2}). Here, \(P_{\rm pair}\) represents the usual contribution from the pair potential in the virial term. Additionally, \(P_{\rm act}\) is the extra contribution arising from the coupling between velocity and vector potential. The exact contribution of this term is determined by the form of \(\vec{A}_i\). In the results section, we present the simulation results for the pressure.

{\bf Specific Heat.} To establish that the proposed Nos\'e-Poincar\'e thermostat generates a correct canonical distribution, one may compare the specific heat $(C)$ of the system in both canonical and microcanonical ensembles. In the canonical ensemble, $C^{NVT}$ is directly related to the fluctuation in the total energy of the system, and the specific heat per particle can be written as
\begin{equation}
    C^{NVT} = \frac{\beta^2}{N}\left< (\delta \mathcal{H})^2 \right>
    \label{C_NVT}
\end{equation}
On the other hand, $\left< (\delta \mathcal{H})^2 \right> = - \frac{\partial \left< \mathcal{H} \right>}{\partial \beta}$, and thus one can determine the specific heat by computing the slope of the average total energy as a function of temperature,
\begin{equation}
    C^{NVT}_{\rm slope} = - \left. \frac{\partial \langle E \rangle}{\partial T} \right|_{V,N}
    \label{C_NVT_slope}
\end{equation}
where $\langle E \rangle$ is the average total energy of the system.
Similarly, in a microcanonical ensemble with fixed energy $E$, the mean-square fluctuations of the potential energy $E_{\mathrm{pot}}$ are given by $\left\langle (\delta E_{\mathrm{pot}})^2 \right\rangle = \frac{3 N}{2 \beta^2}\left( 1 - \frac{3}{2C} \right)$ \cite{lebowitz1967ensemble}. Hence, the specific heat per particle is
\begin{equation}
    C^{NVE} =  \frac{3}{2\left( 1 - \frac{2\beta^2}{3}\frac{\left< (\delta E_{\mathrm{pot}})^2 \right>}{N} \right)}
    \label{C_NVE}
\end{equation}
In the results section, we show that the specific heat computed in the canonical ensemble using the proposed Nos\'e-Poincar\'e thermostat correctly reproduces the results obtained in the microcanonical ensemble, thereby reconfirming the correctness of the proposed symplectic integration scheme.
\section{The Symplectic Algorithm \label{sec_algorithm}}
The idea of numerically solving Hamilton's equations of motion is to sequentially propagate the Hamiltonian system over small time steps $\delta t$. One can show that the propagation of the system over a sufficiently small $\delta t$ can be interpreted as a canonical transformation (or a symplectic map) of the phase space coordinates, thus preserving the symplectic structure of the Hamiltonian system. In a symplectic algorithm, one propagates the system over a short time $\delta t$ via a time evolution operator that approximates the exact time evolution operator up to some power of $\delta t$ (typically it is $\delta t^2$). Since the exact time evolution operator of a Hamiltonian system is given by the operator ${\rm e}^{t \mathcal{L}}$, with $\mathcal{L}$ the Liouville operator, one refers to methods for deriving symplectic algorithms as Liouville operator formalism.

Now we consider the Nos\'e-Poincar\'e equations of motion and put all the phase space variables of this system in a $(6N+2)$ dimensional vector, $\vec{\mu} = \left( \mathcal{R}, \mathcal{P}^{\, \prime}, \Theta, \Omega^{\, \prime}, s, p_s\right)^T$. Then, the exact time evolution of these coordinates to a state $\mu(\delta t)$, starting from an intial state $\mu(0)$, can be formally expressed as

\begin{equation}
    \vec{\mu}(\delta t) = {\rm e}^{\delta t \mathcal{L}} \vec{\mu}(0)
\label{eq_tevol}
\end{equation}
with the Liouville operator $\mathcal{L}$, defined by
\begin{eqnarray}
\mathcal{L} &=& \sum_{i=1}^N \biggl(\frac{\partial \widetilde{\mathcal{H}}}{\partial \vec{p}_i^{\, \prime}}\cdot \frac{\partial}{\partial \vec{r}_i} + \frac{\partial \widetilde{\mathcal{H}}}{\partial \omega_i^{\, \prime}}\cdot \frac{\partial}{\partial \theta_i} + \frac{\partial \widetilde{\mathcal{H}}}{\partial p_s}\cdot \frac{\partial}{\partial s} \nonumber \\
&-& \frac{\partial \widetilde{\mathcal{H}}}{\partial \vec{r}_i}\cdot \frac{\partial}{\partial \vec{p}_i^{\, \prime}} - \frac{\partial \widetilde{\mathcal{H}}}{\partial \theta_i}\cdot \frac{\partial}{\partial \omega_i^{\, \prime}} - \frac{\partial \widetilde{\mathcal{H}}}{\partial s}\cdot \frac{\partial}{\partial p_s}\biggr)\\
&=& \mathcal{L}_r + \mathcal{L}_\theta + \mathcal{L}_s + \mathcal{L}_{p^{\, \prime}} + \mathcal{L}_{\omega^{\, \prime}} + \mathcal{L}_{p_s}. \nonumber
\label{full_liouville_operator}
\end{eqnarray}
The exact time evolution according to Eq.~(\ref{eq_tevol}) can in general not be evaluated, even for the case of an infinitesimal time step $\delta t$. However, for small $\delta t$ an approximate form of the time evolution operator in terms of a product of exponential subpropagators can be found for which the problem can be solved. The latter approximation is based on the following formula for non-commuting operators $A$ and $B$ \cite{Yoshida1990, Forest1990}:
\begin{equation}
    {\rm e}^{\delta t (A+B)} = \prod_{i=1}^k {\rm e}^{a_i \delta t A} {\rm e}^{b_i \delta t B} + \mathcal{O}((\delta t)^{n+1}) \, , \label{eq_split1}
\end{equation}
where the coefficients $a_i$ and $b_i$ can be derived in a systematic manner, as shown by Suzuki \cite{Suzuki1992}. For $n=2$ and $k=2$, one finds \cite{Yoshida1990} $a_1 = a_2 = \frac{1}{2}$, $b_1 = 1$, and $b_2 = 0$ and thus
\begin{equation}
    {\rm e}^{\delta t (A+B)} = {\rm e}^{\frac{1}{2} \delta t A} {\rm e}^{\delta t B} {\rm e}^{\frac{1}{2} \delta t A} + \mathcal{O}((\delta t)^{3}) \, . \label{eq_split2} 
\end{equation}
In the following, we use Eq.~(\ref{eq_split2}) as a basis for the derivation of a second-order symplectic algorithm.

Before we perform the splitting of the time evolution operator, we first evaluate the action of operators ${\rm e}^{\delta t \mathcal{L}_r}$, ${\rm e}^{\delta t \mathcal{L}_p}$, etc.~on phase-space coordinates.

The operator ${\rm e}^{\delta t \mathcal{L}_r}$ only affects the particle's coordinates $\mathcal{R}$. Thus, by representing the exponential by its series expansion and using Eq.~(\ref{np_eom_dr}), one obtains
\begin{eqnarray}
   && {\rm e}^{\delta t \mathcal{L}_r \delta t} \mathcal{R}(0) \nonumber \\
   && = \sum_{n=0}^\infty \sum_{i=1}^N \frac{\delta t^n}{n!}\left(\frac{\vec{p}_i^{\, \prime}(0)-s(0) \vec{A}_i(0)}{ms(0)} \cdot \frac{\partial}{\partial \vec{r}_i} \right)^n \mathcal{R}(0).
   \label{eq_tevolr}
\end{eqnarray}
For the case $\vec{A}_i = 0$, Eq.~(\ref{eq_tevolr}) implies a simple translation of the coordinates, e.g.~for particle $i$ one gets $\vec{r}_i(\delta t)=\vec{r}_i(0) + \frac{\vec{p}^{\, \prime}_i(0)}{ms(0)} \delta t$. However, if one introduces vector potentials that depend on the particle's coordinates one obtains in general a non-linear time evolution from $t=0$ to $t=\delta t$. Similar conclusions hold for the operators  ${\rm e}^{\delta t \mathcal{L}_\theta}$, ${\rm e}^{\delta t \mathcal{L}_{p^{\prime}}}$, and ${\rm e}^{\delta t \mathcal{L}_{\omega^{\prime}}}$. Below, we will discuss these operators for a specific choice of the $\vec{A}_i$'s.

As we shall see now, the time evolution that results from the operators ${\rm e}^{\delta t \mathcal{L}_s}$ and ${\rm e}^{\delta t \mathcal{L}_{p_s}}$ has a non-linear form. Using Eq.~(\ref{np_eom_s}), we can write
\begin{eqnarray}
    {\rm e}^{\delta t \mathcal{L}_s} s(0) &\equiv& s(\delta t) \nonumber \\
    &=&\sum_{n=0}^\infty \sum_{i=1}^N \frac{\delta t^n}{n!}\left(s(0)\frac{p_s(0)}{Q} \frac{\partial}{\partial s} \right)^n s(0) \nonumber \\
    &=& s(0) \exp\left( \frac{p_s(0)}{Q} \delta t \right).
    \label{time_evolution_of_s}
\end{eqnarray}
Thus, the time-scaling variable $s$ has an exponential time update.

The next step is to consider ${\rm e}^{\delta t \mathcal{L}_{p_s}}$. For convenience, we write Eqs.~(\ref{np_eom_dps}) and (\ref{delta_H}) for $\dot{p}_s$ as $\dot{p}_s\left(\mathcal{R}, \mathcal{P}^{\, \prime}, \Theta, \Omega^{\, \prime}, s, p_s \right) = \mathcal{M}\left( \mathcal{R}, \mathcal{P}^{\, \prime}, \Theta, \Omega^{\, \prime}, s \right) - \frac{p_s^2}{2Q}$. Then, the time evolution of $p_s$ can be written as
\begin{eqnarray}
    {\rm e}^{\delta t \mathcal{L}_{p_s}} p_s(0) &\equiv& p_s(\delta t) \nonumber \\
    &=& \sum_{n=0}^\infty \sum_{i=1}^N \frac{\delta t^n}{n!}\left(\left( \mathcal{M}_0-\frac{p_s^2(0)}{2Q} \right) \frac{\partial}{\partial p_s} \right)^n p_s(0) \nonumber\\
    = p_s(0) &+& \delta t \left( \mathcal{M} - \frac{p_s^2}{2Q} \right) - \frac{\delta t^2}{2!}\frac{p_s}{Q}\left( \mathcal{M} - \frac{p_s^2}{2Q} \right) \nonumber \\
    + \frac{\delta t^3}{3!Q}&&\hspace{-0.45cm}\left( \mathcal{M} -\frac{p_s^2}{2Q} \right)\left( \frac{p_s^2}{Q} -\left(\mathcal{M}-\frac{p_s^2}{2Q}\right) \right)+\dots.
    \label{time_evolution_of_ps}
\end{eqnarray}
where $\mathcal{M}_0$ corresponds to $\mathcal{M}$ at time $t=0$. From Eq.~(\ref{time_evolution_of_ps}), it is not obvious whether there is a closed form for $p_s(\delta t)$. To circumvent this problem, we split the operator $\mathcal{L}_{p_s}$ as $\mathcal{L}_{p_s} = \mathcal{L}_{p_s}^{(1)} + \mathcal{L}_{p_s}^{(2)}$ with $\mathcal{L}_{p_s}^{(1)} = \mathcal{M} \frac{\partial}{\partial p_s}$ (note that $\mathcal{M}$ does not depend on $p_s$!) and $\mathcal{L}_{p_s}^{(2)} = -\frac{p_s^2}{2 Q}$. Then, we apply the time-evolution operators ${\rm e}^{\delta t \mathcal{L}_{p_s}^{(1})}$ and ${\rm e}^{\delta t \mathcal{L}_{p_s}^{(2)}}$ separately on $p_s$.

With respect to $\mathcal{L}_{p_s}^{(1)}$, we obtain
\begin{eqnarray}
    {\rm e}^{\delta t \mathcal{L}_{p_s}^{(1)}} p_s(0) &=& \sum_{n=0}^\infty \sum_{i=1}^N \frac{\delta t^n}{n!}\left(\mathcal{M} \frac{\partial}{\partial p_s} \right)^n p_s(0) \nonumber \\
    &=& p_s(0) + \delta t \mathcal{M}, 
    \label{time_evolution_of_ps1}
\end{eqnarray}
and with respect to $\mathcal{L}_{p_s}^{(2)}$
\begin{eqnarray}
    {\rm e}^{\delta t \mathcal{L}_{p_s}^{(2)}} p_s(0) &=& \sum_{n=0}^\infty \sum_{i=1}^N \frac{\delta t^n}{n!}\left(-\frac{p_s^2(0)}{2Q} \frac{\partial}{\partial p_s} \right)^n p_s(0) \nonumber \\
    &=& p_s(0) \sum_{i=1}^N \left(- \frac{\delta t}{2Q} \; p_s(0) \right)^n \nonumber \\
    &=& p_s(0)\left[ \frac{1}{1+\frac{\delta t}{2Q} \, p_s(0)}  \right] . \label{eq_explps2}
    \label{time_evolution_of_ps2}
\end{eqnarray}
Thus, the operator ${\rm e}^{\delta t \mathcal{L}_{p_s}^{(2)}}$ is associated with a nonlinear time evolution of $p_s$.

Using a second-order scheme according to Eq.~(\ref{eq_split2}), we propose the following splitting scheme of the time evolution operator ${\rm e}^{\delta t \mathcal{L}}$:
\begin{eqnarray}
    {\rm e}^{\delta t \mathcal{L}} &\thickapprox& {\rm e}^{\frac{1}{2} \delta t \mathcal{L}_{\omega^{\prime}}} {\rm e}^{\frac{1}{2} \delta t \mathcal{L}_{p^{\prime}}} {\rm e}^{\frac{1}{2} \delta t\mathcal{L}_{p_s}^{(1)}} {\rm e}^{\frac{1}{2} \delta t \mathcal{L}_{p_s}^{(2)}} \nonumber \\
    &&{\rm e}^{\frac{1}{2} \delta t \mathcal{L}_s} {\rm e}^{\frac{1}{2} \delta t \mathcal{L}_\theta} {\rm e}^{\delta t \mathcal{L}_r } {\rm e}^{\frac{1}{2} \delta t \mathcal{L}_\theta} {\rm e}^{\frac{1}{2} \delta t \mathcal{L}_s} \nonumber \\
    &&{\rm e}^{\frac{1}{2} \delta t \mathcal{L}_{p_s}^{(2)}} {\rm e}^{\frac{1}{2} \delta t \mathcal{L}_{p_s}^{(1)}} {\rm e}^{\frac{1}{2} \delta t \mathcal{L}_{p^{\prime}}} {\rm e}^{\frac{1}{2} \delta t \mathcal{L}_{\omega^{\prime}}} \, .
    \label{eq_split_the_teo}
\end{eqnarray}
Here, the order of the application of operators is optimized such that the resulting scheme is computationally most efficient with respect to the number of force calculations. 

To obtain the time evolution from an initial state $\mu(0)$ at $t=0$ to a state $\mu(\delta t)$ at $t=\delta t$, the product of subpropagators, Eq.~(\ref{eq_split_the_teo}), is sequentially applied to $\mu(0)$. Of course, the details of the updates of the phase space variables depend on the choice of the vector potential $\vec{A}_i$ for each particle $i$. We set 
\begin{equation}
\vec{A}_i = K \vec{S}_i \, ,
\label{eq_casiulis}
\end{equation}
as proposed by Casiulis {\it et al.}~\cite{Casiulis2020_1, Casiulis2020_2}. However, in what follows, we also discuss how one can perform the updates for a general choice of the vector potentials $\vec{A}_i$, depending in some manner on the coordinates of the particles and their spins. The algorithm that we obtain according to Eq.~(\ref{eq_split_the_teo}) is as follows:
\begin{enumerate}
\item[I.] First, we have to apply ${\rm e}^{\frac{1}{2}\delta t \mathcal{L}_{\omega^{\prime}}}$ on $\Omega^{\, \prime}(0)$ and thus for particle $i$ we get
\begin{equation}
\omega^{\, \prime}_i\left(\frac{\delta t}{2}\right) 
= \omega_i^{\prime}(0) + \frac{\delta t}{2} \, \dot{\omega}_i^{\prime}(0) \, ,
   \label{1st_step}
\end{equation}   
where $\dot{\omega}_i^{\prime}(0)$ is given by Eq.~(\ref{np_eom_domega}). Since $\dot{\omega}_i^{\prime}(0)$ does not depend on $\omega_i^{\, \prime}$, one can independently update the $\omega^{\, \prime}_i$'s for each particle $i$ according to Eq.~(\ref{1st_step}) to obtain $\Omega^{\, \prime}(\delta t/2)$.
\item[II.] Next, the operator ${\rm e}^{\frac{1}{2}\delta t \mathcal{L}_{p^{\prime}}}$ acts on $\mathcal{P}^{\, \prime}(0)$. For $\vec{A}_i = K \vec{S}_i$ the derivatives of $\vec{A}_i$ with respect to the coordinates in Eqs.~(\ref{np_eom_dpx}) and (\ref{np_eom_dpy}) vanish and thus $\dot{p}_i^{\, \prime x}$ and $\dot{p}_i^{\, \prime y}$ do not depend on ${p}_i^{\, \prime x}$ or ${p}_i^{\, \prime y}$. Therefore, the update of the momentum $\vec{p}^{\, \prime}_i$ for particle $i$ is
\begin{equation}
\vec{p}^{\, \prime}_i\left(\frac{\delta t}{2}\right) = \vec{p}^{\, \prime}_i(0) + \frac{\delta t}{2} \, \dot{\vec{p}}^{\, \prime}_i(0) \, .
\label{2nd_step}
\end{equation}
Applying this equation independently to each particle $i$, we obtain $\mathcal{P}^{\, \prime}\left( \frac{\delta t}{2} \right)$.
\item[III.] We have defined the operator $\mathrm{e}^{\frac{1}{2} \delta t \mathcal{L}^{(1)}_{p_s}}$ via Eq.~(\ref{time_evolution_of_ps1}). It leads to the following intermediate update of the $s$ momentum, that we denote by $\widetilde{p}_s$,
\begin{eqnarray}
    &&\widetilde{p}_s\left( \frac{\delta t}{2} \right) = p_s (0) + \frac{\delta t}{2} \times \nonumber\\
    &&\times \mathcal{M}\left( \mathcal{R}(0), \mathcal{P}^{\, \prime}\left(\frac{\delta t}{2}\right), \Theta (0), \Omega^{\, \prime}\left(\frac{\delta t}{2}\right), s(0) \right) .
    \label{3rd_step}
\end{eqnarray}
\item[IV.] The application of $\mathrm{e}^{\frac{1}{2} \delta t \mathcal{L}^{(2)}_{p_s}}$ on $\widetilde{p}_s(\delta t/2)$ using Eq.~(\ref{time_evolution_of_ps2}) completes the update of $p_s$ at $\delta t/2$,
\begin{equation}
    p_s\left( \frac{\delta t}{2} \right) = \widetilde{p}_s\left(\frac{\delta t}{2}\right)\left[\frac{1}{1+\frac{\delta t}{4Q} \, \widetilde{p}_s\left(\frac{\delta t}{2}\right)}\right] .
    \label{4th_step}
\end{equation}
\item[V.] The application of $\mathrm{e}^{\frac{\delta t}{2} \mathcal{L}_s}$ on $s(0)$ updates the scaling variable $s$ at half time step, 
%
\begin{equation}
    s\left( \frac{\delta t}{2} \right) = s(0) \, \exp\left(\frac{\delta t}{2Q} \; p_s\left( \frac{\delta t}{2} \right)\right) ,
    \label{5th_step}
\end{equation}
see Eq.~(\ref{time_evolution_of_s}).
\item[VI.] Since $\dot{\omega}^{\prime}_i$ does not depend on $\omega_i$, the half-time update of $\theta_i$ applying $\mathrm{e}^{\frac{\delta t}{2} \mathcal{L}_\theta}$ is straightforward,
\begin{equation}
    \theta_i\left( \frac{\delta t}{2}\right) = \theta_i(0) + \frac{\delta t}{2Is(\delta t/2)} \, \omega^{\prime}_i\left( \frac{\delta t}{2}\right) .
    \label{6th_step}
\end{equation}
\item[VII.] Now the update of the coordinates, $\mathcal{R}$, at the full time step $\delta t$ can be performed via the operator ${\rm e}^{\delta t \mathcal{L}_r}$. For particle $i$, one obtains
\begin{equation}
    \vec{r}_i\left( \delta t\right) = \vec{r}_i(0) + \frac{\delta t}{m} \left[ \frac{\vec{p}^{\, \prime}_i \left(\frac{\delta t}{2}\right)}{s\left(\frac{\delta t}{2}\right)} - K\vec{S}_i\left(\frac{\delta t}{2}\right) \right] .
    \label{7th_step}
\end{equation}
The new coordinates $\mathcal{R}$ have to be used below to obtain $p_s$, $\vec{p}^{\, \prime}_i$, and $\omega^{\prime}_i$ at time $\delta t$.

\item[VIII.] Next, for each particle $i$ we compute $\theta_i(\delta t)$ analogously to Eq.~(\ref{6th_step}),
\begin{equation}
    \theta_i\left(\delta t\right) = \theta_i\left(\frac{\delta t}{2}\right) + \frac{\delta t}{2 I s(\delta t/2)} \, \omega^{\prime}_i\left( \frac{\delta t}{2}\right) ,
    \label{8th_step}
\end{equation}
to obtain $\Theta(\delta t)$.
\item[IX.] The update of the scaling variable $s$ at full time step is similar to Eq.~(\ref{5th_step}), 
\begin{equation}
    s\left( \delta t \right) = s\left(\frac{\delta t}{2}\right) \exp\left(\frac{\delta t}{2Q} \; p_s\left( \frac{\delta t}{2} \right)\right) .
    \label{9th_step}
\end{equation}
\item[X.] Similar to Eq.~(\ref{4th_step}),  
\begin{equation}
    \widetilde{p}_s\left( \delta t \right) = p_s\left( \frac{\delta t}{2} \right)\left[\frac{1}{1+\frac{\delta t}{4Q} \, p_s \left(\frac{\delta t}{2}\right)}\right] ,
    \label{10th_step}
\end{equation}
\item[XI.] And similar to Eq.~(\ref{3rd_step}), we obtain
\begin{eqnarray}
    &&p_s\left( \delta t \right) = \widetilde{p}_s \left( \delta t\right) + \frac{\delta t}{2} \times \nonumber \\
    &&\times \mathcal{M}\left( \mathcal{R}\left( \delta t\right), \mathcal{P}^{\, \prime}\left(\frac{\delta t}{2}\right), \Theta (\delta t), \Omega^{\, \prime}\left(\frac{\delta t}{2}\right), s(\delta t) \right)\-.
    \label{11th_step}
\end{eqnarray}
\item[XII.] Then, the update of the momentum for a particle $i$ is given by
\begin{equation}
\vec{p}^{\, \prime}_i\left(\delta t\right) = \vec{p}^{\, \prime}_i\left(\frac{\delta t}{2}\right) + \frac{\delta t}{2} \, \dot{\vec{p}}^{\, \prime}_i\left(\delta t\right) \, ,
   \label{12th_step}
\end{equation}
where we use $\mathcal{R}(\delta t)$, $\Theta(\delta t)$, and $s(\delta t)$ to compute $\dot{\vec{p}}^{\, \prime}_i\left(\delta t\right)$ according to Eqs.~(\ref{np_eom_dpx}) and (\ref{np_eom_dpy}). The update for all the particles gives $\mathcal{P}^{\, \prime}(\delta t)$.
\item[XIII.] Finally, the step
\begin{equation}
\omega^{\, \prime}_i\left(\delta t\right) 
= \omega_i^{\prime}\left(\frac{\delta t}{2}\right) + \frac{\delta t}{2} \, \dot{\omega}_i^{\prime}\left(\delta t\right) \, ,
   \label{13th_step}
\end{equation}
for all particles $i$ yields $\Omega^{\, \prime}(\delta t)$ and completes the update of all the phase space variables.
\end{enumerate}
Steps I.~to XIII.~describe a symplectic algorithm for the case that the vector potentials $\vec{A}_i$ ($i=1, \dots, N$) do not depend on the coordinates $\mathcal{R}$ and depend on the spin variables such that $\vec{A}_i = K \vec{S}_i$. For the general case that the $\vec{A}_i$'s are a function of $\mathcal{R}$ and a general function of $\Theta$ (such that $\vec{A}_i$ also depends on spins $\vec{S}_j$ with $j\neq i$), in particular steps I., II,, XII., and XIII.~are more complicated because then one has $[p_i^{\prime \alpha}, p_j^{\prime \beta}] \neq 0$ and $[\omega_i^{\, \prime}, \omega_j^{\, \prime}] \neq 0$ for $i\neq j$ and $\alpha, \beta = x, y$, i.e.~the translational and rotational momenta do not commute with each other. For this more general case, symplectic algorithms are presented elsewhere.

\section{Numerical Simulations \label{sec_simulation}}
Now we present the results of numerical simulations based on the Hamiltonian (\ref{np_hamiltonian}), using the algorithm presented in the previous section. As we shall see, at a given number density $\rho$, the system shows a change of behavior from low to high temperatures. This change occurs around a critical temperature $T_c$. While at low temperature particle clusters with a coherent flock-like motion form, at high temperature the particles are in a gas phase. In the following, we analyze how the energy is distributed among the different degrees of freedom and we ratioanlize the definition of temperature, as given by Eq.~(\ref{eq_temperature}), is correct. Furthermore, the distributions of particle velocities, $V_\alpha$ with $\alpha = x, y$, are investigated. We find that the average particle velocity, $\langle v_\alpha \rangle$, exhibits a bifurcation at low temperature, where, towards low temperature, $\langle v_\alpha \rangle$ changes from zero to a finite value.

{\bf Details of the simulation.} We consider two-dimensional systems of $N$ particles where $N$ ranges from 200 to 2000. Periodic boundary conditions are applied in the two spatial directions. For the pair potential $u(r_{ij})$, we use a Weeks-Chandler-Andersen (WCA) potential, given by
\begin{equation}
u(r_{ij})= 4 \varepsilon\left[\left(\frac{\sigma}{r_{ij}}\right)^{12}-\left(\frac{\sigma}{r_{ij}}\right)^{6}\right]+\varepsilon
\label{WCA}
\end{equation}
for $r_{ij}< 2^{1/6}\sigma$ and $u(r_{ij}) = 0$ otherwise. The energy parameter and the ``diameter'' are respectively set to $\varepsilon = 1$ and $\sigma = 1$ in all simulations. For the function $g(r_{ij}$ in Eq.~(\ref{np_hamiltonian}), we choose
\begin{equation}
g(r_{ij}) = \frac{(r_{ij}-r_c)^4}{h+(r_{ij}-r_c)^4}
\label{gr}
\end{equation}
for $r_{ij}< r_c = 1.5\sigma$ and $g(r_{ij})=0$ otherwise. In Eq.~(\ref{gr}), we set $h = 0.00001$ to make $g(r_{ij})$ almost a step function of radial distance. For the vector potential, we consider $\vec{A} = K \vec{S}_i$ with $K=1.0$. The masses are set to $Q=1.0$ and $m=1.0$. Time is measured in units of $\tau = \sqrt{m \sigma^2/\varepsilon}$. The Nos\'e-Poincar\'e equations of motion, described in detail in the previous section, are integrated with a time step $\delta t$ varying from $0.0005$ to $0.005$.

\begin{figure}
\centering
\includegraphics[width=0.425\textwidth]{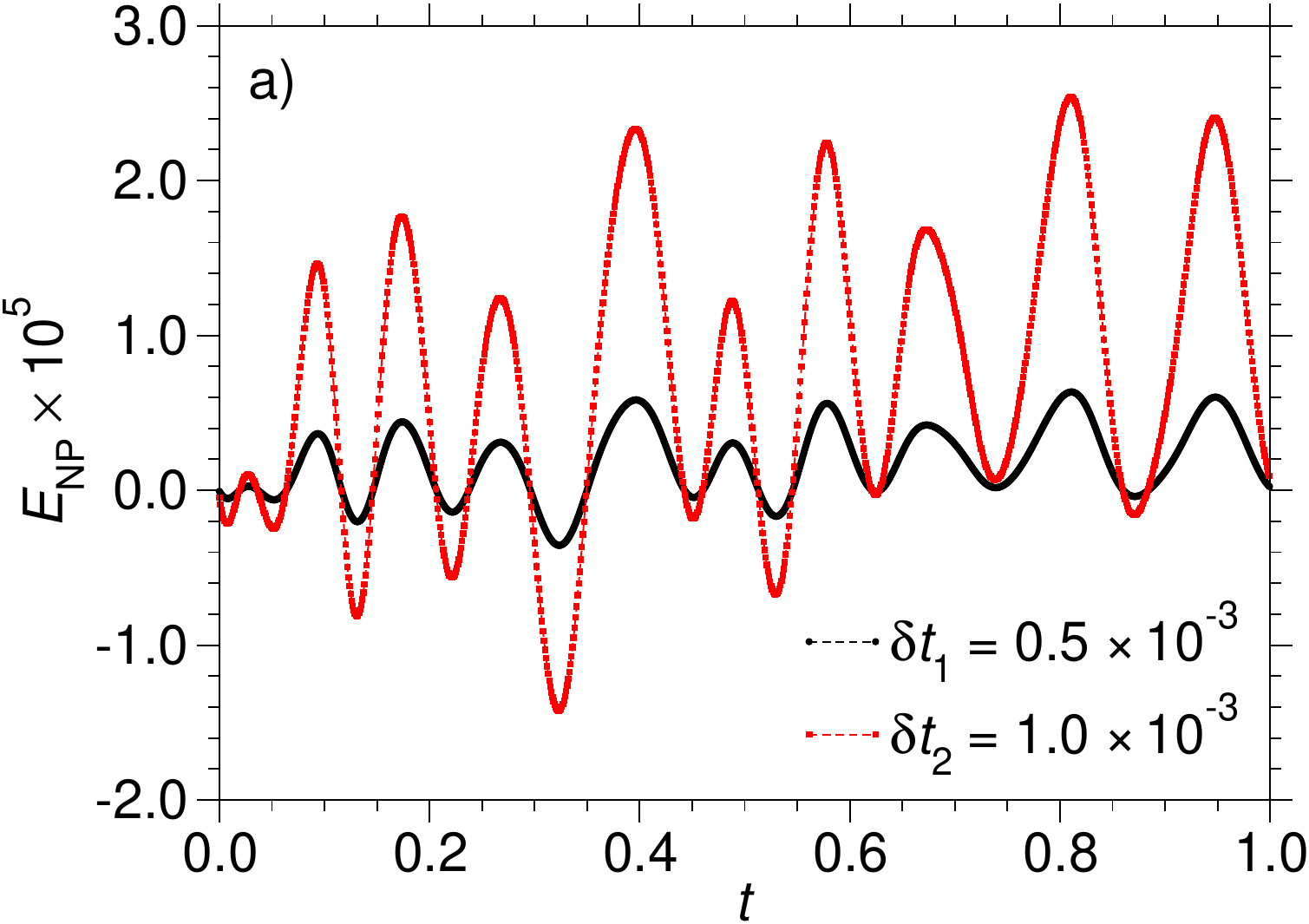}
\includegraphics[width=0.45\textwidth]{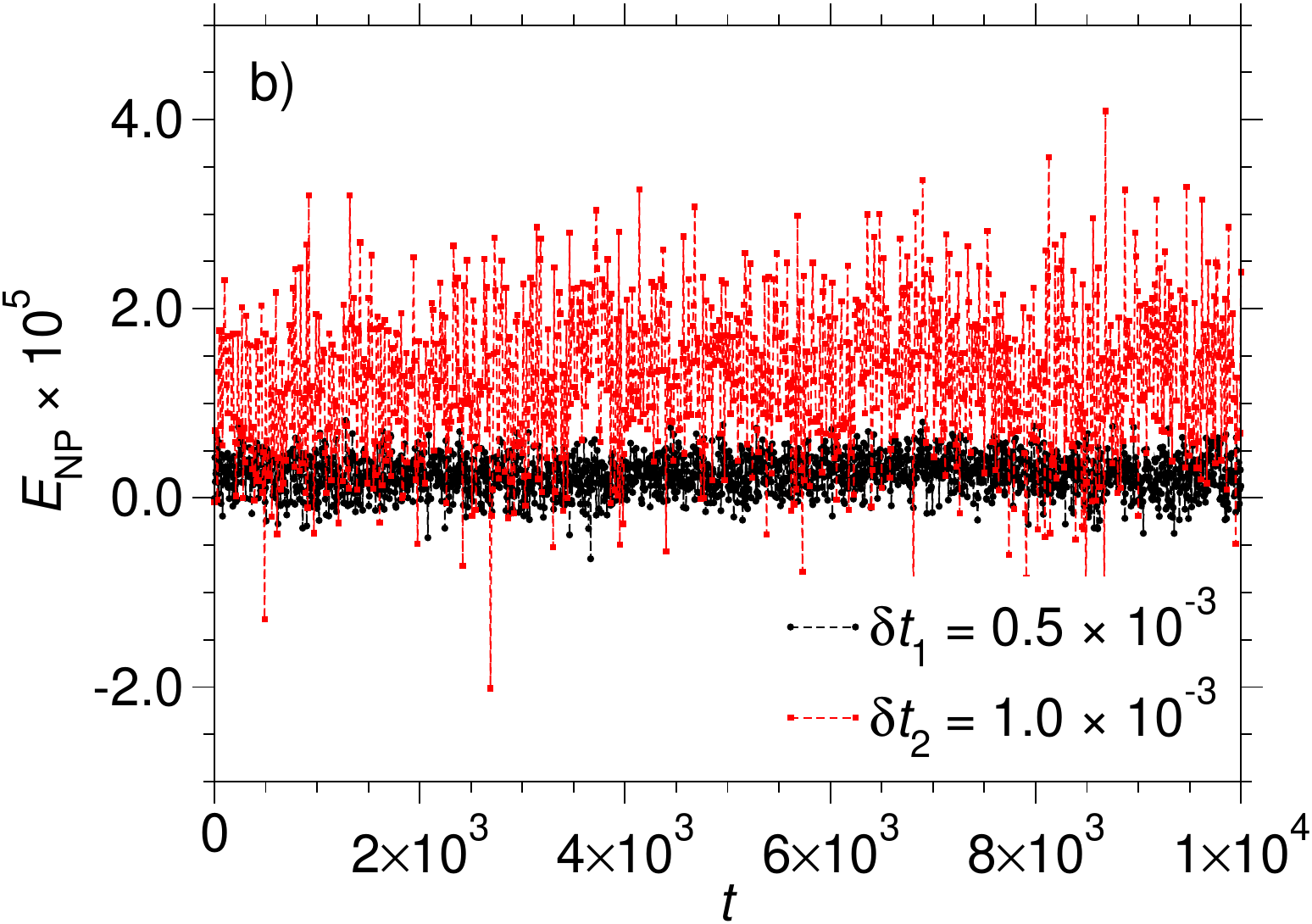}
\caption{Test of the energy conservation: Time series of total energy of a Nos\'e-Poincar\'e system with $N=128$ particles starting from the same equilibrated initial conditions with the different time steps $\delta t_1 = 0.5 \times 10^{-3}$ (red) and $\delta t_2 = 1.0 \times 10^{-3}$ (black). Here, we consider $T=0.15$ and $\rho = 0.2$. a) Energy time series for short times, $0\le t \le 1.0$ and b) for long times, $0 \le t \le 10^4$. \label{fig1}}
\end{figure}
\begin{figure*}
\centering
\includegraphics[width=0.41\textwidth]{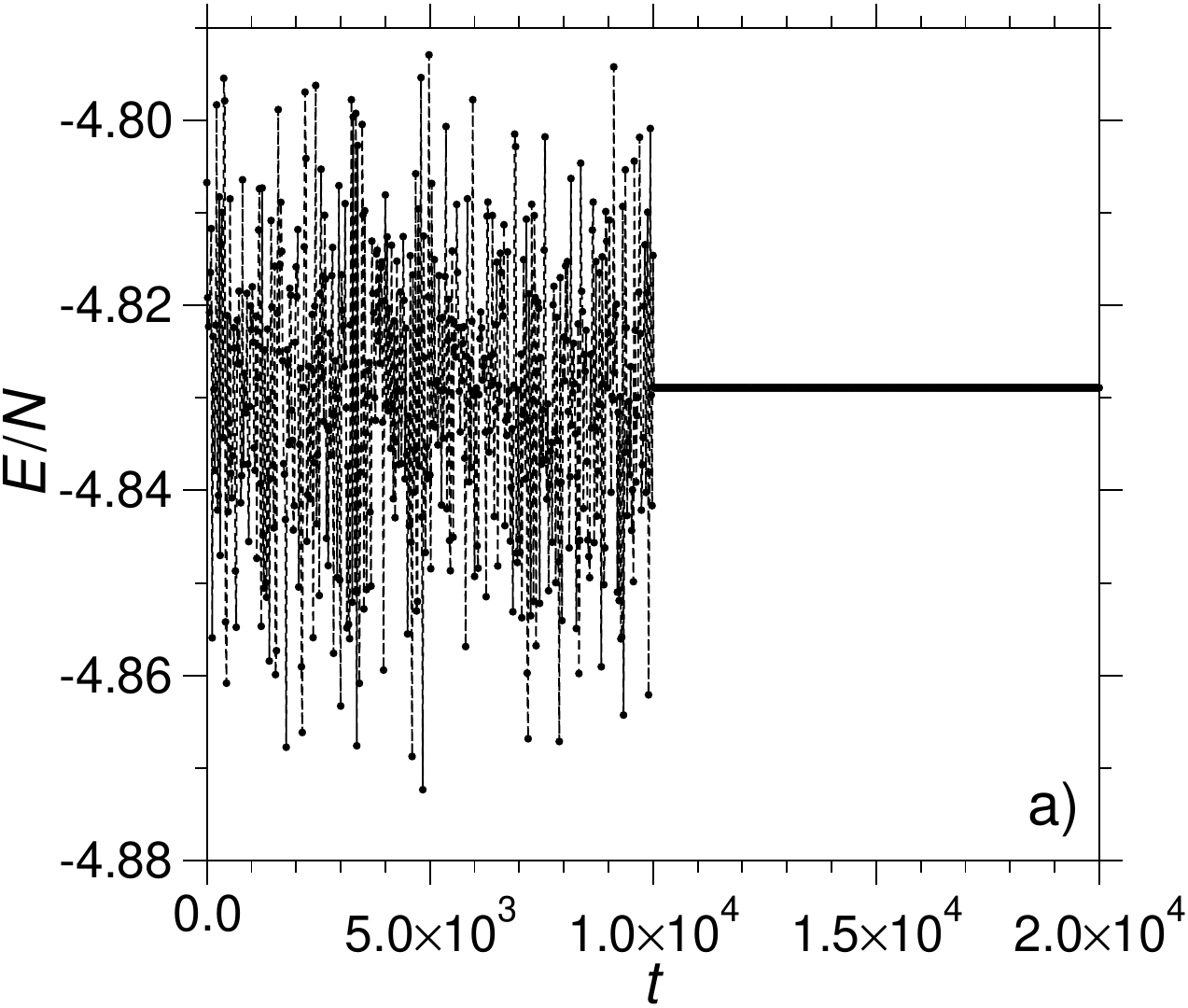}
\includegraphics[width=0.41\textwidth]{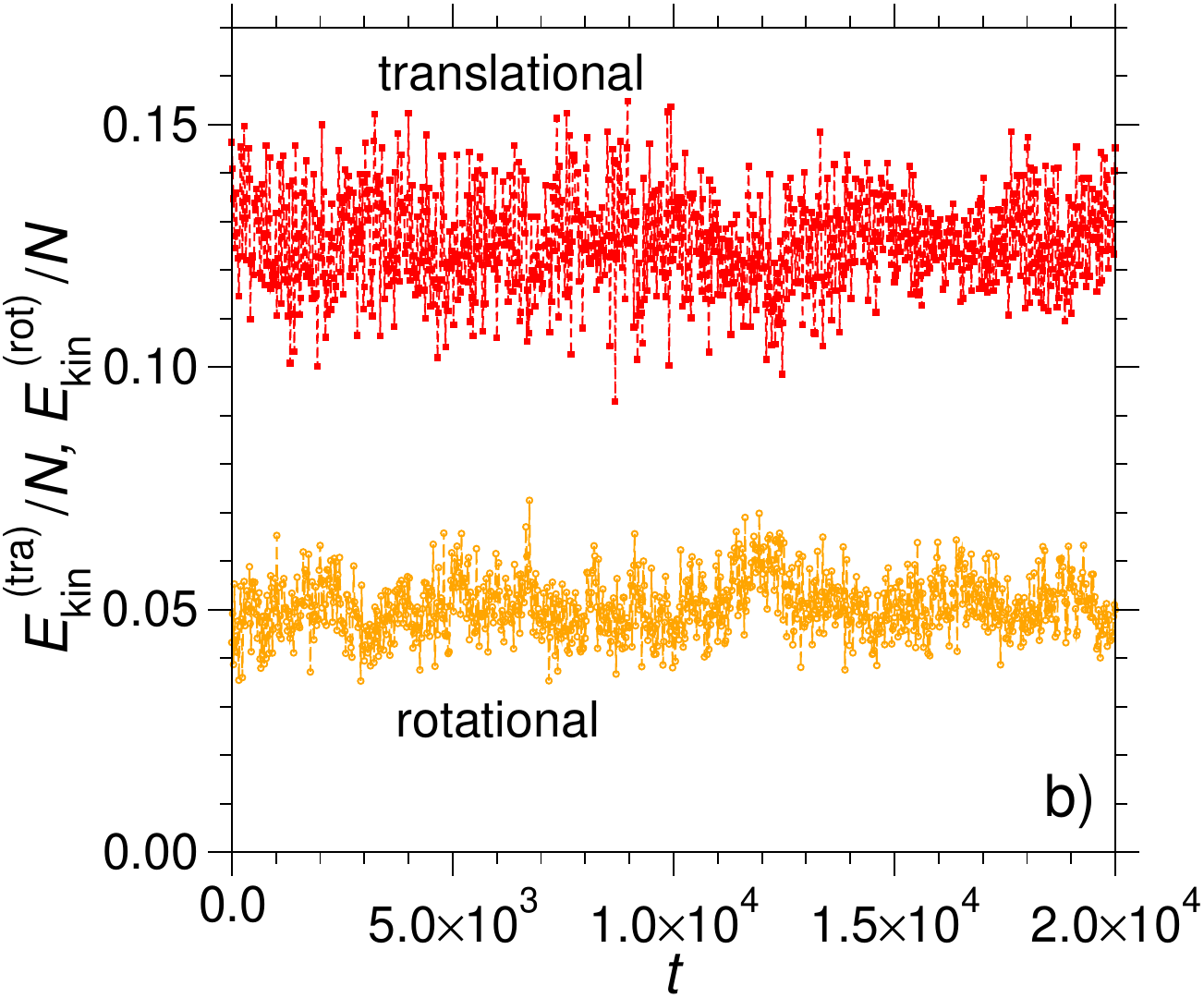}
\includegraphics[width=0.42\textwidth]{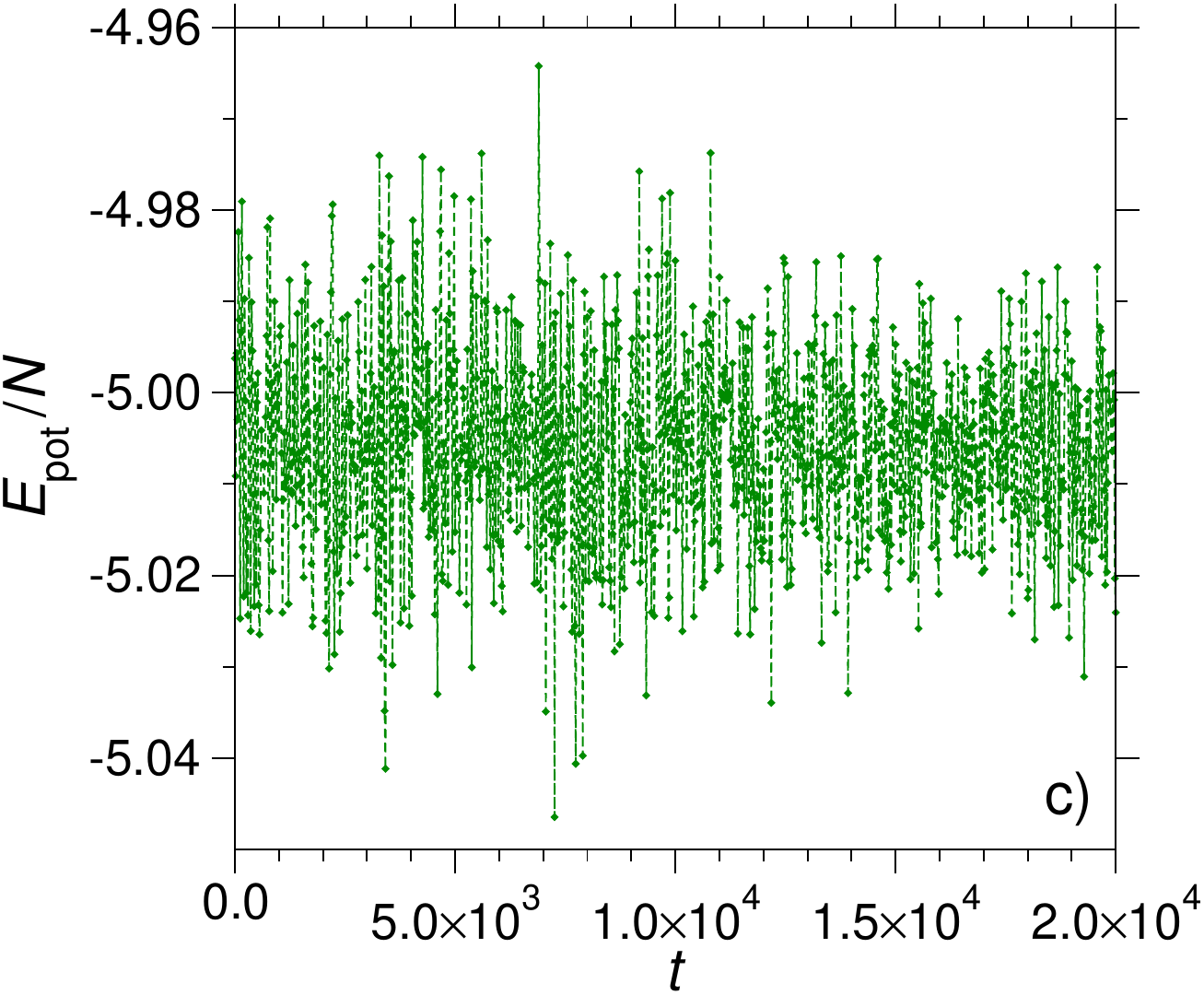}
\includegraphics[width=0.4\textwidth]{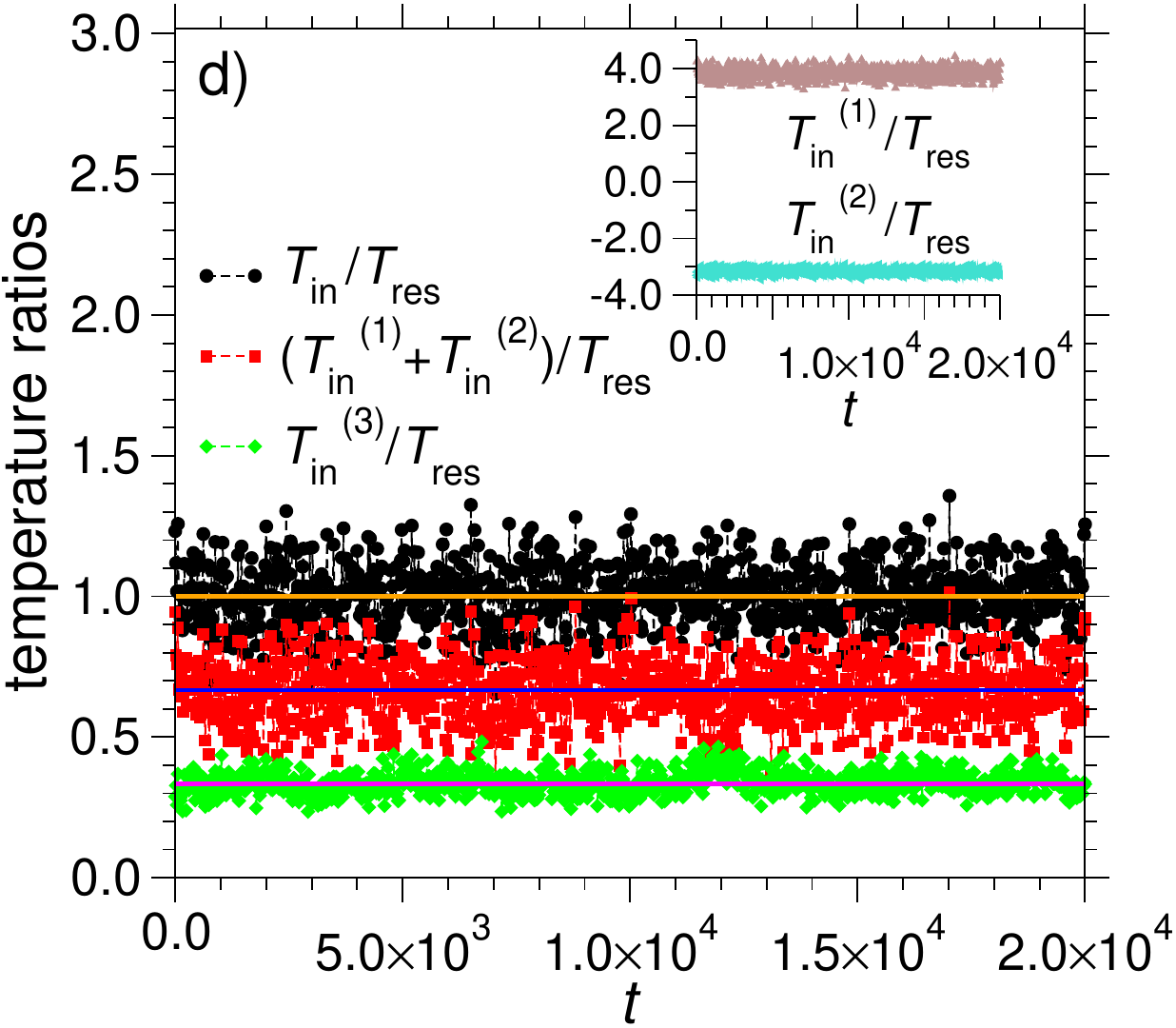}
\caption{Time series of energies and various contributions to temperature for a fluid with $N = 128$, $T = 0.1$, $K=1.0$, $\rho=0.2$, $Q = 1.0$. The time step for the integration of the equations of motion is $\delta t = 0.001$. a) Total energy per particle, $E/N$, b) translational and rotational kinetic energy, $E_{\mathrm{kin}}^{\mathrm{(tra)}}/N$ and $E_{\mathrm{kin}}^{\mathrm{(rot)}}/N$, respectively, c) potential energy per particle, $E_{\mathrm{pot}}/N$, and d) various ``temperature'' ratios (see text). \label{fig2}}
\end{figure*}
\begin{figure}
\centering
\includegraphics[width=0.42\textwidth]{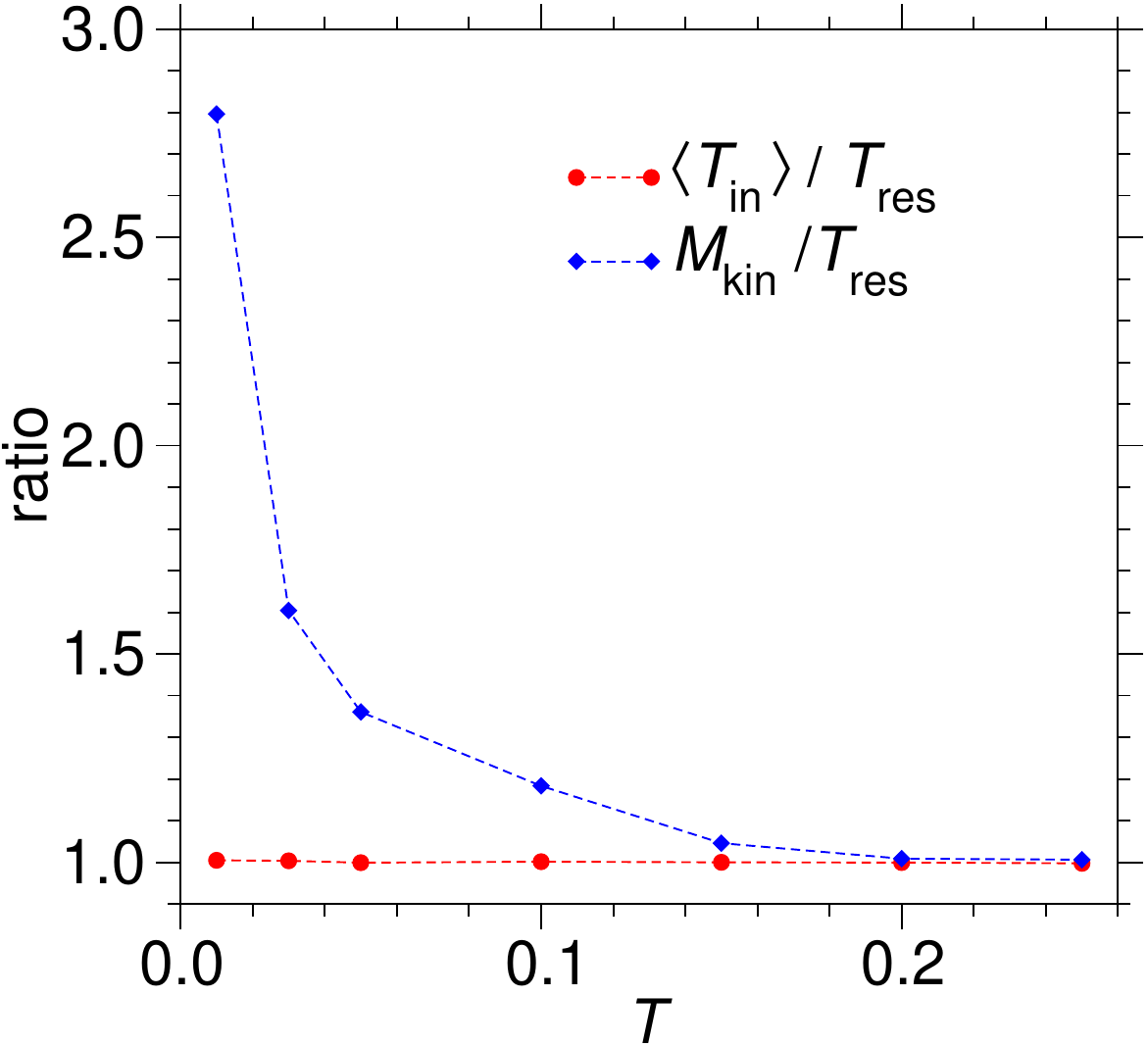}
\caption{Ratios $\langle T_{\mathrm{in}} \rangle/T_{\mathrm{red}}$ and $M_{\mathrm{kin}}/T_{\mathrm{res}}$ as a function of temperature. \label{fig3}}
\end{figure}
The symplectic algorithm, presented in Sec.~\ref{sec_algorithm}, is a second-order scheme. Thus, the discretization error of the total energy, $E_{\rm NP}$, should scale with the time step squared, $\delta t^2$. That this indeed the case, is demonstrated in Fig.~\ref{fig1}. This figure shows time series of the total energy per particle, $E_{\rm NP}/N$, for a system of $N=128$ particles at the density $\rho = 0.2$. Starting with the same initial configuration, the two time series in Fig.~\ref{fig1} correspond to the time steps $\delta t_1 = 0.5 \times 10^{-3}$ (red curve) and $\delta_2 = 1.0 \times 10^{-3}$ (black curve). For short times, $0\le t \le 1.0$, we can infer from Fig.~\ref{fig1}a) that the amplitude of the oscillations for $\delta t_2$ are a factor of 4 higher than those for $\delta t_1$, as required for $\delta t_2/\delta t_1 = 2$. Moreover, for both time steps the oscillations are in phase. This is no longer the case for long times, but, as indicated by Fig.~\ref{fig1}b for $t\le 10^4$, the energy is also conserved on long time scales, maintaining a stable amplitude of the descretization error.

{\bf Canonical Sampling.} To further demonstrate that our symplectic algorithm leads to a correct canonical sampling, we show in Fig.~\ref{fig2} time series of energies and various contributions to temperature, performing a long run over $2\times 10^4\, \tau$ where we switch off the thermostat after $10^4\, \tau$. The total energy per particle, $E/N$, is displayed in Fig.~\ref{fig2}a (here, in the case of the thermostatted system, $E$ corresponds to the energy without the kinetic and potential energy associated with $p_s$ and $s$, respectively). We see that after switching off the thermostat, the energy $E/N$ is constant, as required for the microcanonical ensemble. Figures \ref{fig2}b and \ref{fig2}c show the the corresponding time series of translational and rotational energy per particle that are respectively defined as
\begin{equation}
  \frac{E_{\mathrm{kin}}^{\mathrm{(tra)}}}{N} = \sum_i \frac{((\vec{p}_i - \vec{A}_i)^2}{2m}, \quad 
  \frac{E_{\mathrm{kin}}^{\mathrm{(rot)}}}{N} = \sum_i \frac{\omega_i^2}{2I} \, .
\end{equation}
These quantities do not seem to show any difference when the thermostat is switched off at $t=10^4\,\tau$. This is also true for the potential energy per particle, $E_{\mathrm{pot}}/N$ (Fig.~\ref{fig2}c). The instantaneous temperature, $T_{\mathrm{in}}$, can be splitted into three terms as $T_{\mathrm{in}} = T_{\mathrm{in}}^{(1)} + T_{\mathrm{in}}^{(2)} + T_{\mathrm{in}}^{(3)}$, where the three terms are defined by
\begin{eqnarray}
T_{\mathrm{in}}^{(1)} & = & \frac{1}{3N k_{\mathrm{B}}} \sum_i \frac{\vec{p}_i^{\, 2}}{m} \\
T_{\mathrm{in}}^{(2)} & = & - \frac{1}{3N k_{\mathrm{B}}} \sum_i \frac{\vec{p}_i \cdot \vec{A}_i}{m} \\
T_{\mathrm{in}}^{(3)} & = & \frac{1}{3N k_{\mathrm{B}}} \sum_i \frac{\omega_i^2}{I} \, ,
\end{eqnarray}
cf.~Eq.~(\ref{eq_temperature}). Figure \ref{fig2}d displays the time series of the ratios \( T_{\mathrm{in}}/T_{\mathrm{res}} \), \( (T_{\mathrm{in}}^{(1)} + T_{\mathrm{in}}^{(2)})/T_{\mathrm{res}} \), and \( T_{\mathrm{in}}^{(3)}/T_{\mathrm{res}} \), where \( T_{\mathrm{res}} \) denotes the temperature of the reservoir (for our example, \( T_{\mathrm{res}} = 0.1 \)). We find that the averages of these ratios are given by \( \langle T_{\mathrm{in}}/T_{\mathrm{res}} \rangle = 1.0 \), \( \langle (T_{\mathrm{in}}^{(1)} + T_{\mathrm{in}}^{(2)})/T_{\mathrm{res}} \rangle = 2/3 \), and \( \langle T_{\mathrm{in}}^{(3)}/T_{\mathrm{res}} \rangle = 1/3 \), as required (cf.~the horizontal lines in the figure). The inset shows that the ratios \( T_{\mathrm{in}}^{(1)}/T_{\mathrm{res}} \) and \( T_{\mathrm{in}}^{(2)}/T_{\mathrm{res}} \) are respectively positive and negative numbers with the same order of magnitude. This indicates that, with respect to these terms, the definition of temperature as an average of the kinetic energy cannot be correct. To quantify the error that one introduces by an incorrect definition of temperature in terms of the average kinetic energy, 
\begin{equation}
 M_{\mathrm{kin}} = \left\langle \left( E_{\mathrm{kin}}^{\mathrm{(tra)}}
 + E_{\mathrm{kin}}^{\mathrm{(rot)}} \right) \right\rangle \, ,
 \end{equation}
we now consider the ratio $M_{\mathrm{kin}}/T_{\mathrm{res}}$ as a function of temperature. This is shown in Fig.~\ref{fig3} in comparison to the ratio $\langle T_{\mathrm{in}} \rangle/T_{\mathrm{res}}$ which is of course equal to one for all temperatures. While at high temperature $M_{\mathrm{kin}}$ provides a good estimate of the temperature, towards low temperature the ratio $M_{\mathrm{kin}}/T_{\mathrm{res}}$ deviates more and more from one and reaches almost a value of 3 at the temperature $T=0.01$.


%
\begin{figure}
\centering
\includegraphics[width=0.42\textwidth]{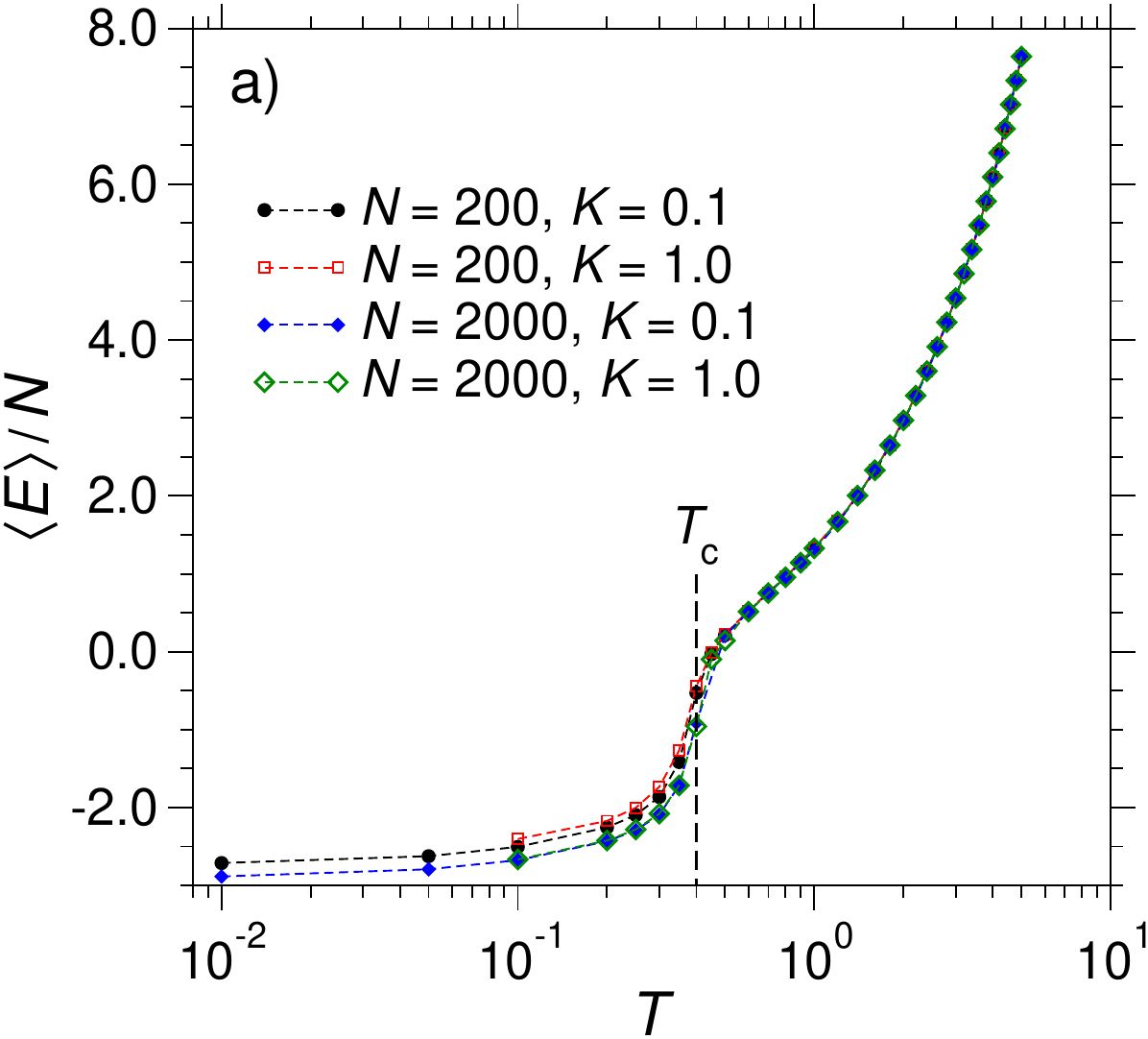}
\includegraphics[width=0.42\textwidth]{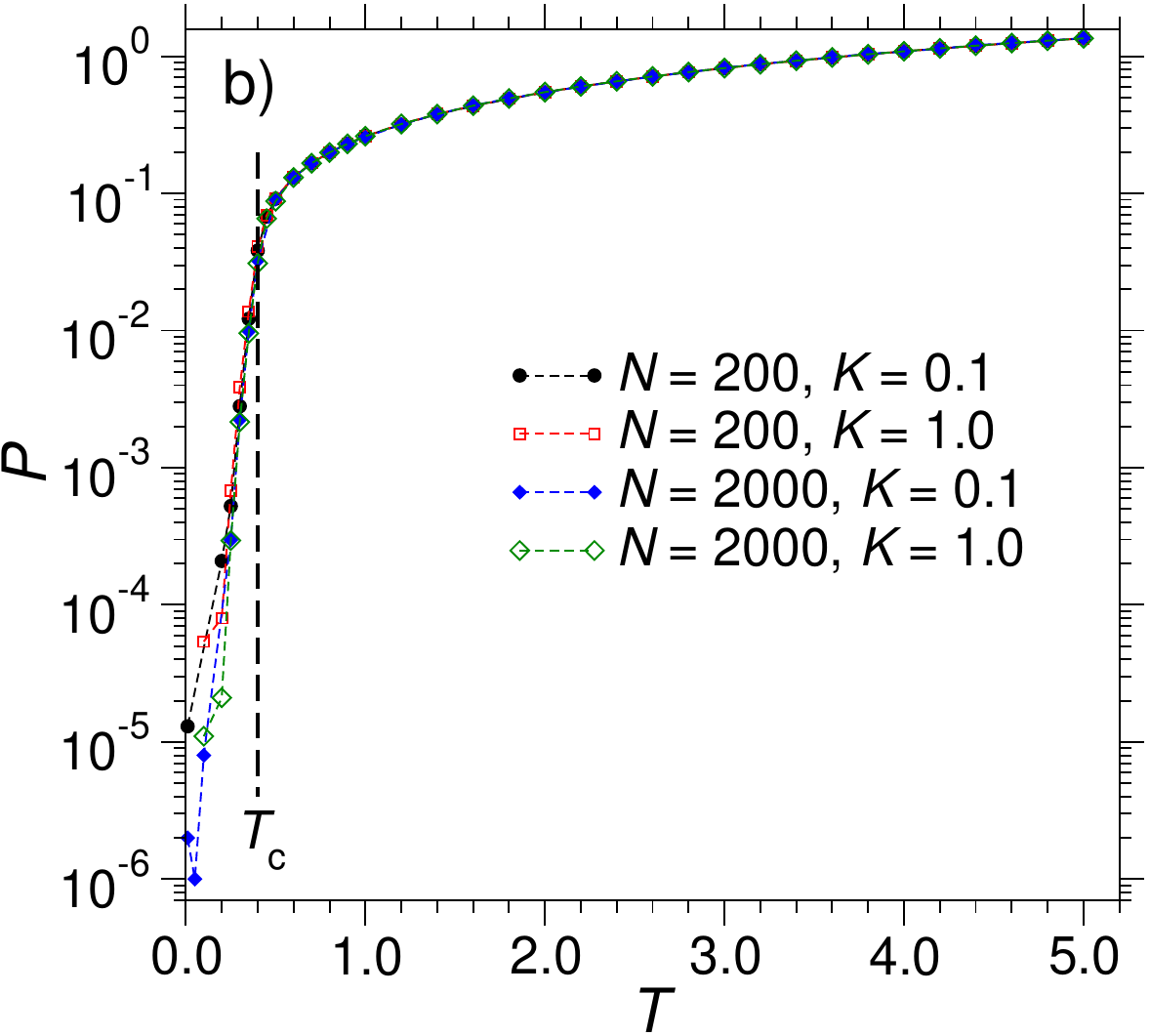}
\caption{a) Mean total energy per particle, $\langle E \rangle/N$, as a function of temperature $T$ for $K=0.1$ and $K=1.0$, both for systems with $N=200$ and $N=2000$ particles. b) The same for the temperature dependence of the pressure $P$. Both in a) and b), the ``critical temperature'' $T_c = 0.4$ is marked by dashed vertical lines. \label{fig4}}
\end{figure}
\begin{figure*}[t]
\centering
\includegraphics[width=\textwidth]{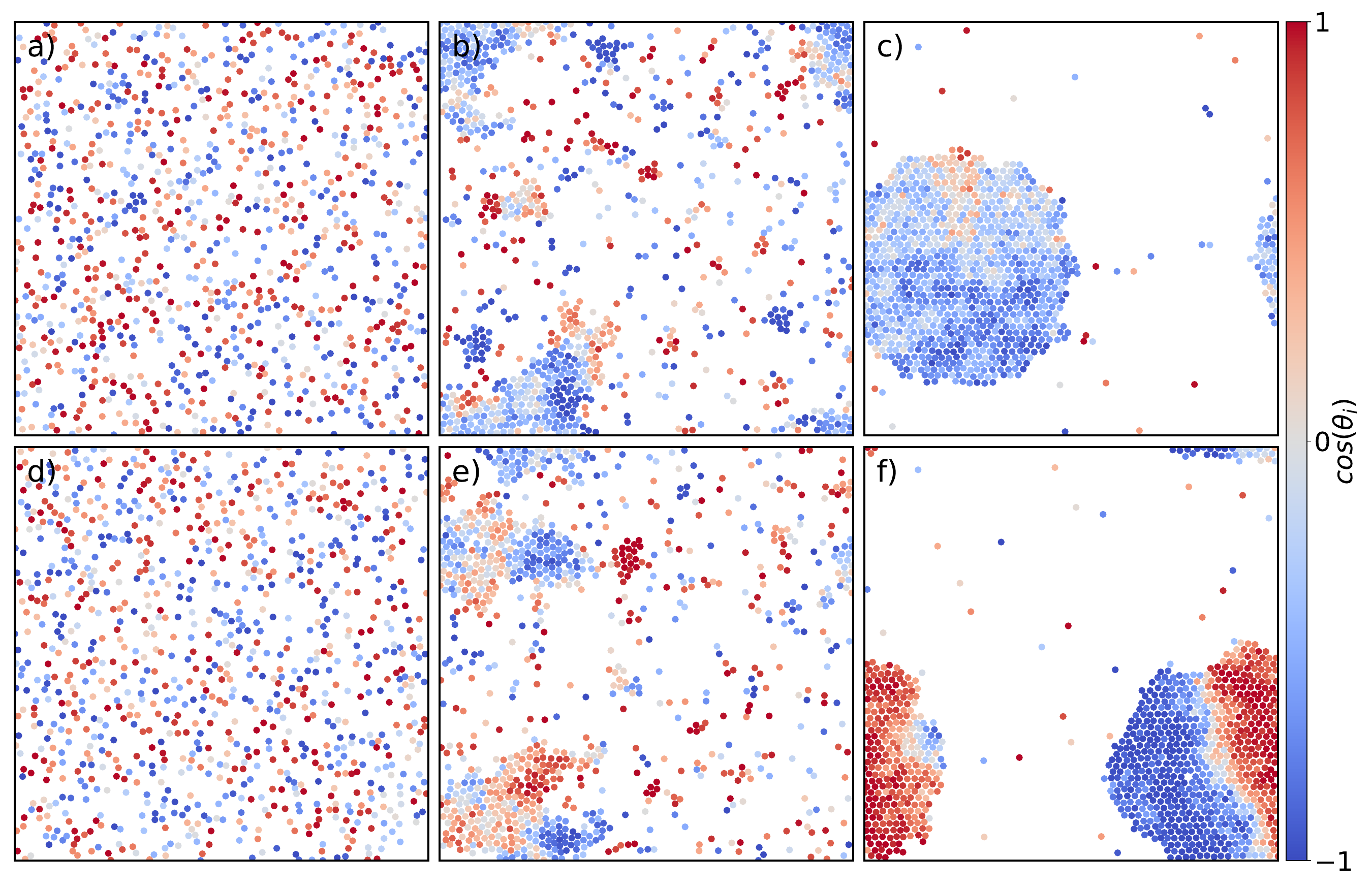}
\caption{Snapshots of systems with $N = 1000$ particles and number density $\rho=0.2$ at a) $T = 1.0$, $K=0.1$, b) $T = 0.4$, $K=0.1$, c) $T = 0.3$, $K=0.1$, d) $T = 1.0$, $K=1.0$, e) $T = 0.4$, $K=1.0$, and f) $T = 0.3$, $K=1.0$. Periodic boundary conditions are applied in $x$ and $y$ direction. The colour map shows the projection of the spin direction of each particle to the $x$ axis, $\cos(\theta_i)$. \label{fig5}}
\end{figure*}

{\bf The fluid-to-cluster transition.} As we shall see now, towards low temperatures the system undergoes a finite-size transition to an ordered cluster phase around a critical temperature $T_c$. However, this transition is different from that recently studied for an $XY$ spin fluid model by Bissinger and Fuchs \cite{Bissinger2023}. In their model, the total interaction potential $u(r) - J g(r) \cos(\theta)$ is always repulsive and therefore, as demonstrated by Bissinger and Fuchs, undergoes Berezinskii-Kosterlitz-Thouless (BKT) transitions \cite{Berezinskii1971, Kosterlitz1974, Kosterlitz2016}. This implies that for an infinitely large system that as a consequence of the Mermin-Wagner theorem the average magnetization always vanishes. This is different from our case where the potential $u(r) - J g(r) \cos(\theta)$ can be attractive and thus for $K=0$ we see a non-vanishing magnetization at low temperatures also for the limit $N\to \infty$, in full agreement with an earlier finding of Casiulis {\it et al.}~\cite{Casiulis2019} for a similar model. Here, our focus is anyway not on the analysis of the BKT scenario, but we investigate the consequences of the velocity-spin coupling for the behavior of the finite-size clusters with a finite magnetization that are observed at low temperature. We shall see that cluster phases with a finite magnetization are associated with a bimodal distribution of the center-of-mass velocity of the particles forming a cluster. 

Figures \ref{fig4}a and \ref{fig4}b show respectively the mean total energy per particle, $\langle E \rangle/N$, and the pressure $P$ as a function of temperature for the two different coupling constants $K=0.1$ and $K=1.0$, in each case for systems of $N=200$ and $N=2000$ particles. In both plots, a ``critical temperature'' $T_c = 0.4$ is marked by dashed vertical lines. Around this temperature, there is the finite-size transition from homogeneous fluid states at high temperature to the low-temperature states where ordered clusters with a finite magnetization form. This transition is reflected in the behavior of the energy per particle that around $T_c$ shows a rapid step-like decrease towards low temperature that becomes slightly sharper with increasing system size, also resulting in a slightly lower energy in the low temperature range for the large system with $N=2000$ particles. At the same time the energy is almost independent of the coupling constant $K$. The pressure exhibits a similar behavior as the energy. Towards low temperature, there is a stress drop around $T_c$ which is slightly sharper for the larger system while the pressure is almost independent on $K$.

Figure \ref{fig5} shows snapshots at different temperatures for systems with $N=1000$ particles and the two values of the coupling constant $K=0.1$ (Fig.~\ref{fig5}a to \ref{fig5}c) and $K=1.0$ (Fig.~\ref{fig5}d to \ref{fig5}f). In these snapshots, the colour on each particle corresponds to its value of the spin direction, projected on the $x$ axis. At $T=1.0$, the system is in a gas phase. Here, the probability of forming even small clusters is very small and there is essentially a random orientation of the spins. This changes around $T_c$ at $T=0.4$. As can be inferred from the corresponding snapshots, at this temperature, there is the formation of small and larger clusters in which the spin orientation is strongly correlated. Eventually, at a temperature below $T_c$, i.e.~at $T=0.3$ in Figs.~\ref{fig5}c and \ref{fig5}f, one observes one big cluster that coexists with a gas phase with a very low density. Here, it is interesting that the cluster for the system with $K=0.1$ exhibits a coherent spin configuration with all spins pointing towards the negative $x$ direction, while the cluster for the system with $K=1.0$ can be split into two parts of opposite spin direction, with an interfacial region between both parts where the spins tend to point to a perpendicular direction, i.e.~in $y$ direction. As a result, the cluster for $K=0.1$ has a larger mean absolute value of the magnetization per particle, $\langle \left| \vec{m} \right| \rangle$ with $\vec{m} = \frac{1}{N} \sum_i \vec{S}_i$, than the one for $K=1.0$. Below we discuss the dependence of $\langle | \vec{m} | \rangle$ on $K$ as well as on system size. We shall see that a finite modulus of the magnetization is associated with a finite center-of-mass velocity of the cluster which is reminiscent of flocking.

\begin{figure}
\centering
\includegraphics[width=0.47\textwidth]{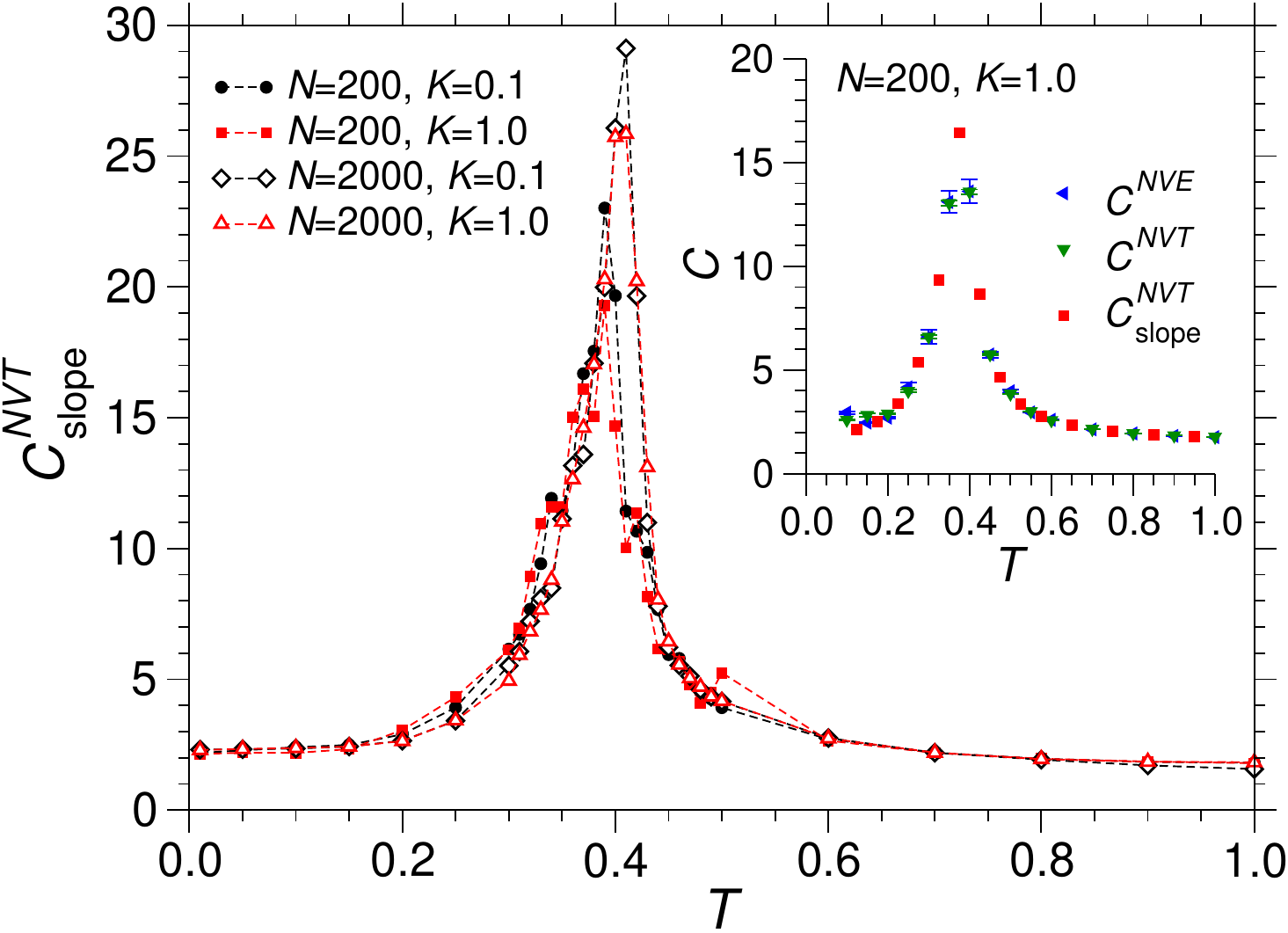}
\caption{Specific heat $C^{NVT}_{\mathrm{slope}}$, as obtained from Eq.~(\ref{C_NVT_slope}), as a function of temperature for $K=0.1$ and $K=1.0$, both for systems with $N=200$ and $N=2000$ particles. The inset shows the specific heat for $N=200$ and $K=1.0$ from the main plot in comparison to the corresponding specific heat in the microcanonical ensemble, $C^{NVE}$, and the specific heat in the canonical ensemble, $C^{NVT}$, as determined from the energy fluctuations. \label{fig6}}
\end{figure}
\begin{figure}
\centering
\includegraphics[width=0.42\textwidth]{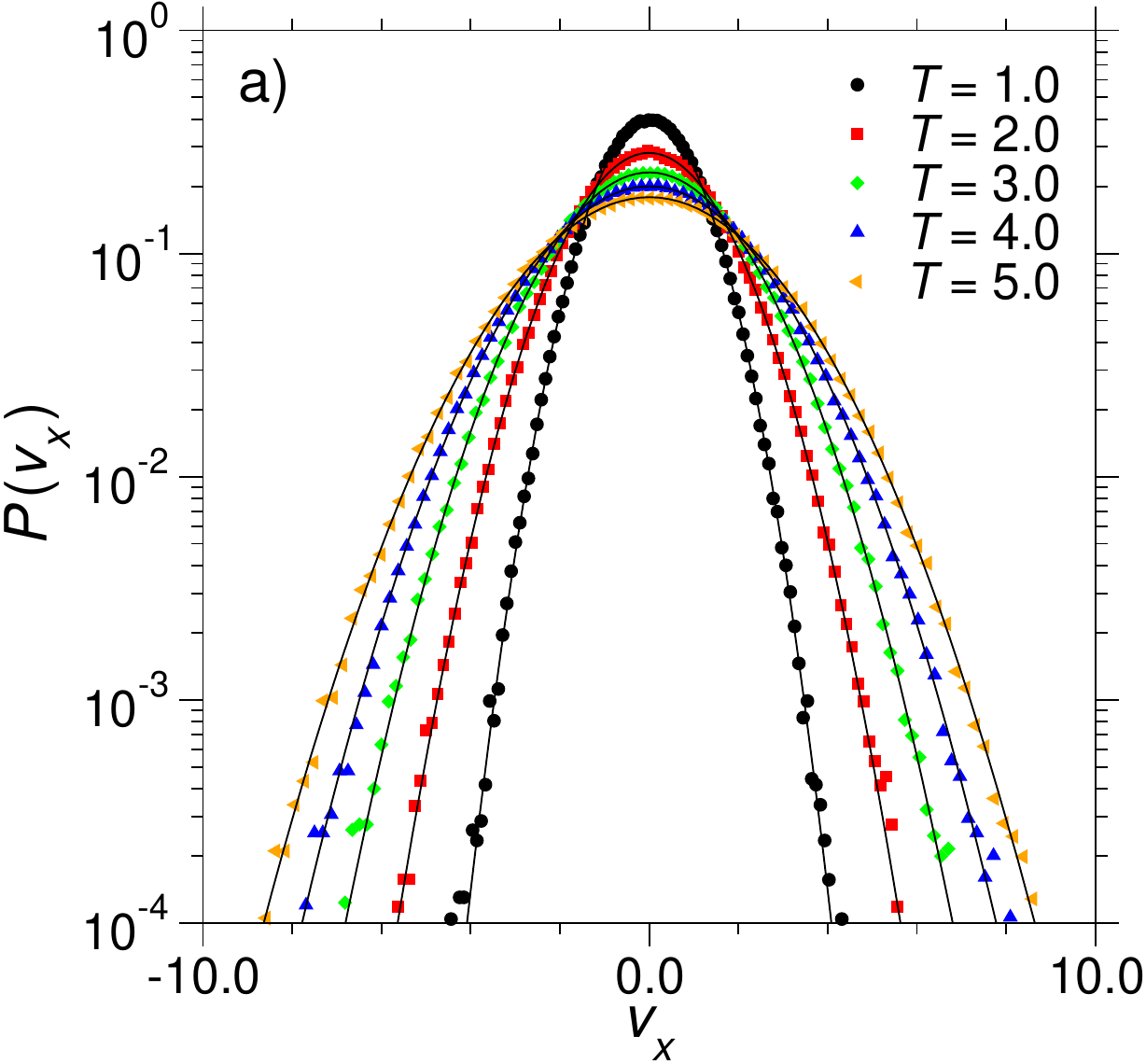}
\includegraphics[width=0.42\textwidth]{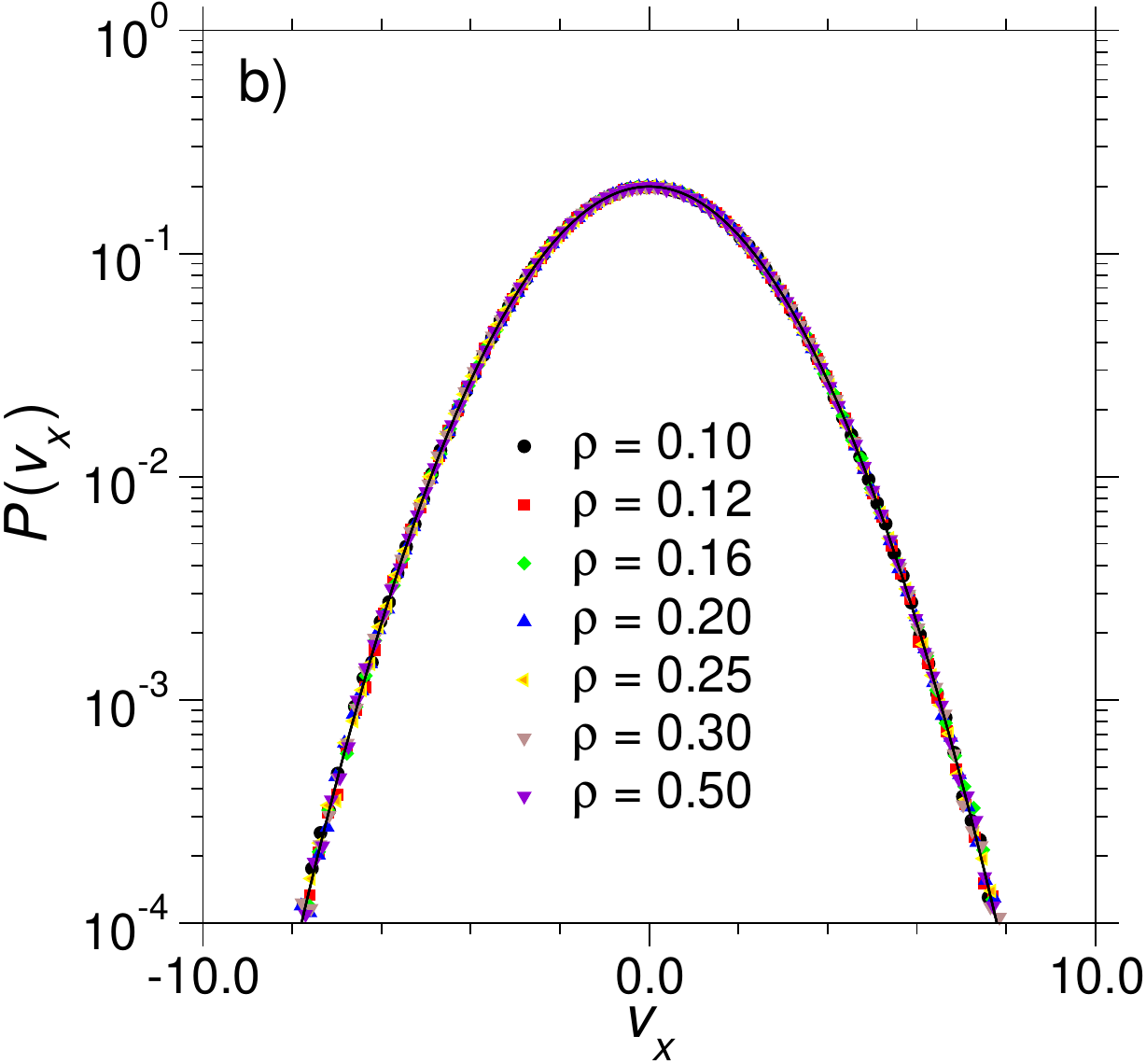}
\caption{a) Distribution of particle velocity in $x$-direction, $P(v_x)$, for high temperatures $5.0 \ge T \ge 1.0$ at the fixed number density $\rho = 0.2$ for systems with $N=2000$ particles. The black straight lines are Gaussian functions, as given by Eq.~(\ref{eq_gaussv}) . b) Distribution $P(v_x)$ for different densities $\rho$ at the fixed temperature $T=4.0$, again using systems with $N=2000$ particles. \label{fig7}}
\end{figure}
\begin{figure}
\centering
\includegraphics[width=0.45\textwidth]{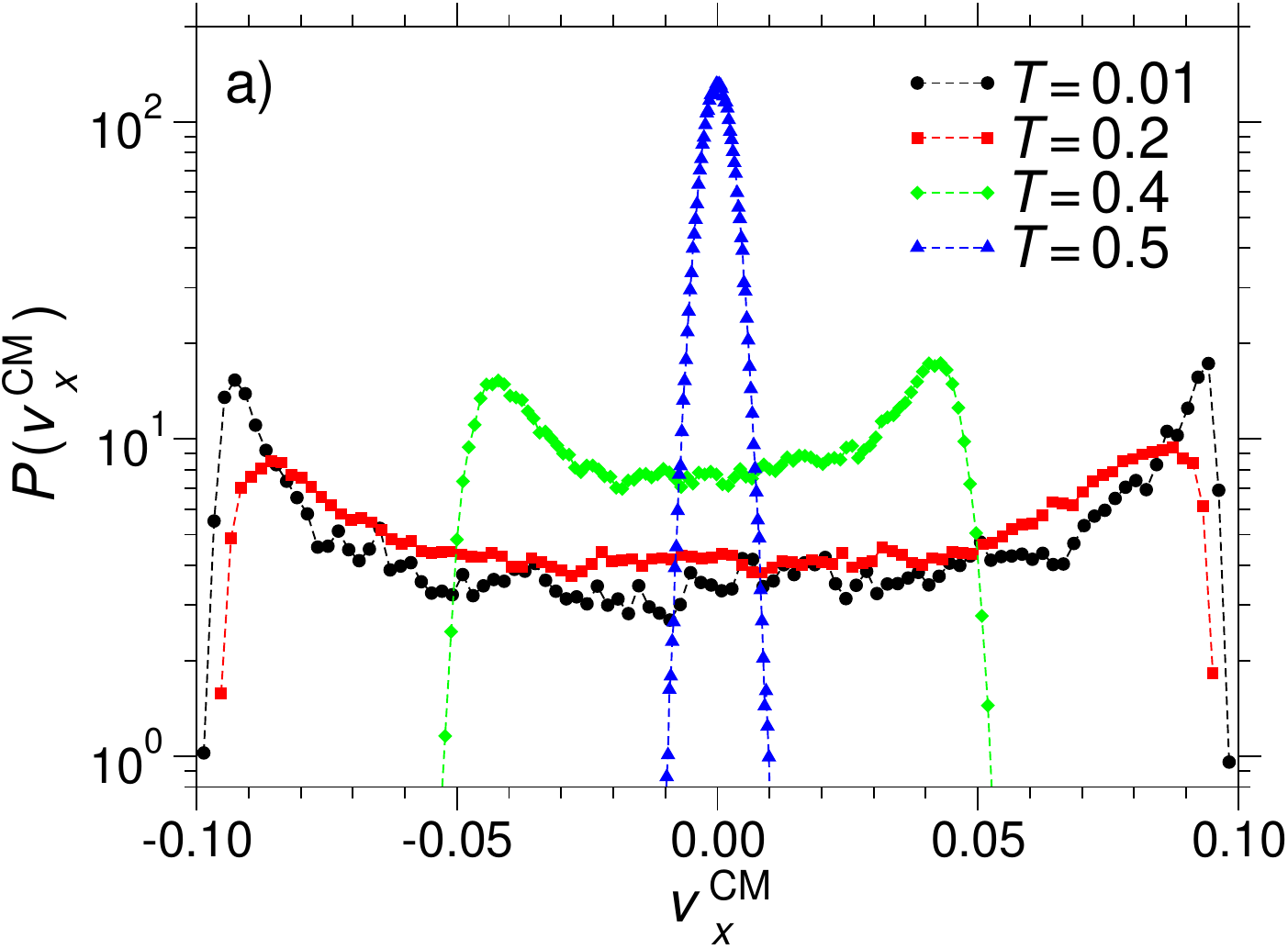}
\includegraphics[width=0.45\textwidth]{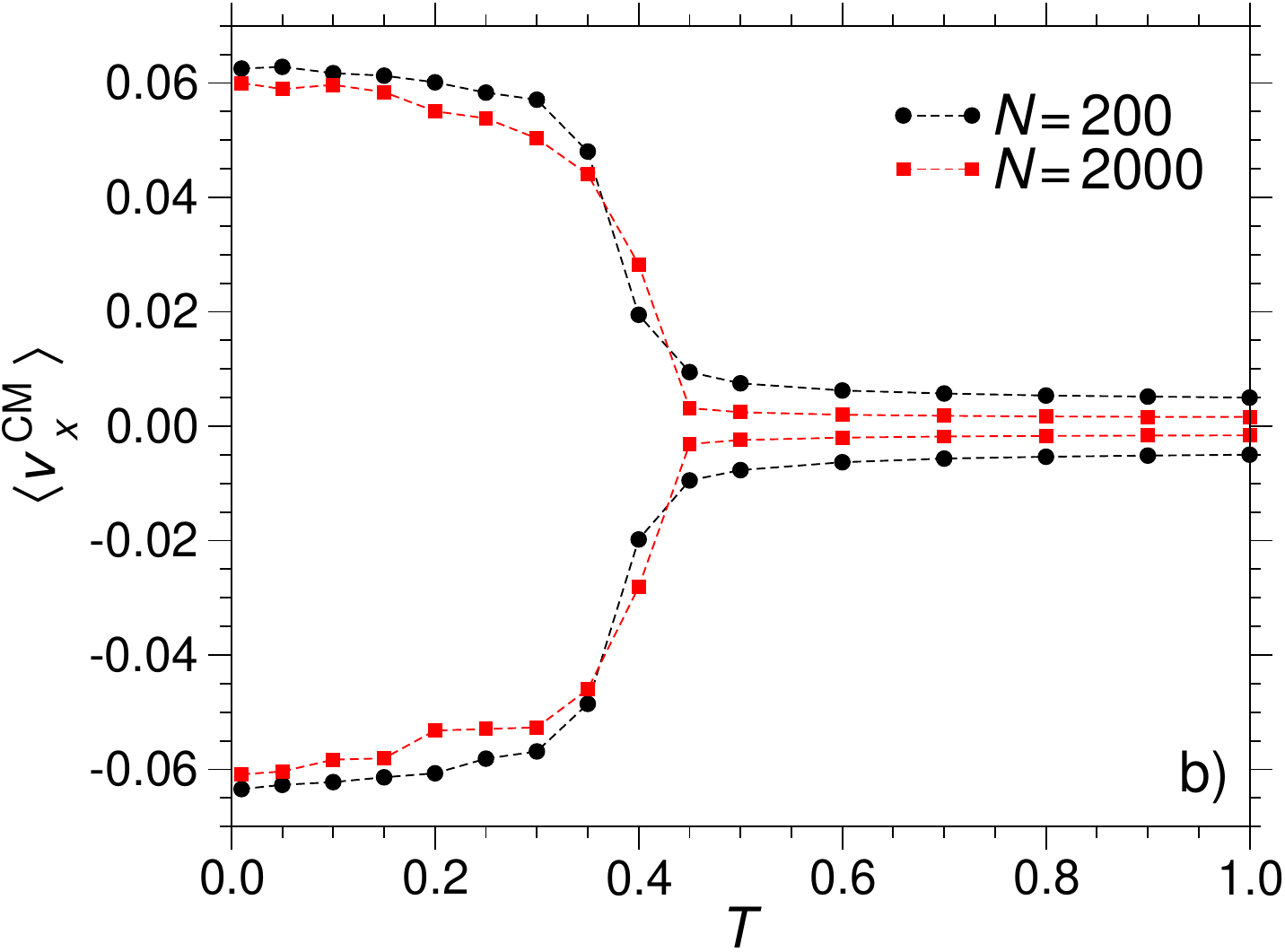}
\caption{a) Distribution of the center-of-mass velocity $v_x^{\mathrm{CM}}$ in $x$ direction at the temperatures $T=0.01$, $T=0.2$, $T=0.4$, and $T=0.5$ for systems with $N=2000$ particles. b) Average center-of-mass velocities $\langle v_x^{\mathrm{CM}} \rangle$, as obtained from the distributions $P(v_x^{\rm CM})$ (see text) for systems with $N=200$ and $N=2000$ particles. \label{fig8}}
\end{figure}

The finite-size fluid-to-cluster transition around the temperature $T_c$ is also reflected in the behavior of the specific heat $C$. The main plot of Fig.~\ref{fig6} displays the temperature dependence of the specific heat $C^{NVT}_{\mathrm{slope}}$, as obtained for the canonical ensemble from the slope of $E(T)$, see Eq.~(\ref{C_NVT_slope}). As for the total energy per particle (Fig.~\ref{fig4}a), this quantity is shown for $K=0.1$ and $K=1.0$ and for each $K$ value for systems with $N=200$ and $N=2000$ particles. The specific heat has a peak around $T_c$. With increasing system size, the height of this peak grows and it slightly shifts to higher temperature. At the same time, there is no significant dependence on the parameter $K$ (note that this is consistent with the behavior of $E/N$, discussed above). These findings do not depend on the way the specific heat is computed as well as on the choice of the ensemble (at least similar results are obtained in the microcanonical ensemble). This is demonstrated in the inset of Fig.~\ref{fig6} for the example of systems with $N=200$ particles and $K=1.0$. Here we see that the canonical specific $C^{NVT}$, as determined via the energy fluctuations via Eq.~(\ref{C_NVT}), as well as the microcanonical specific heat $C^{NVE}$, as computed via Eq.~(\ref{C_NVE}), are in very good agreement with the results for $C^{NVT}_{\mathrm{slope}}$.

\begin{figure*}[t]
\centering
\includegraphics[width=0.45\textwidth]{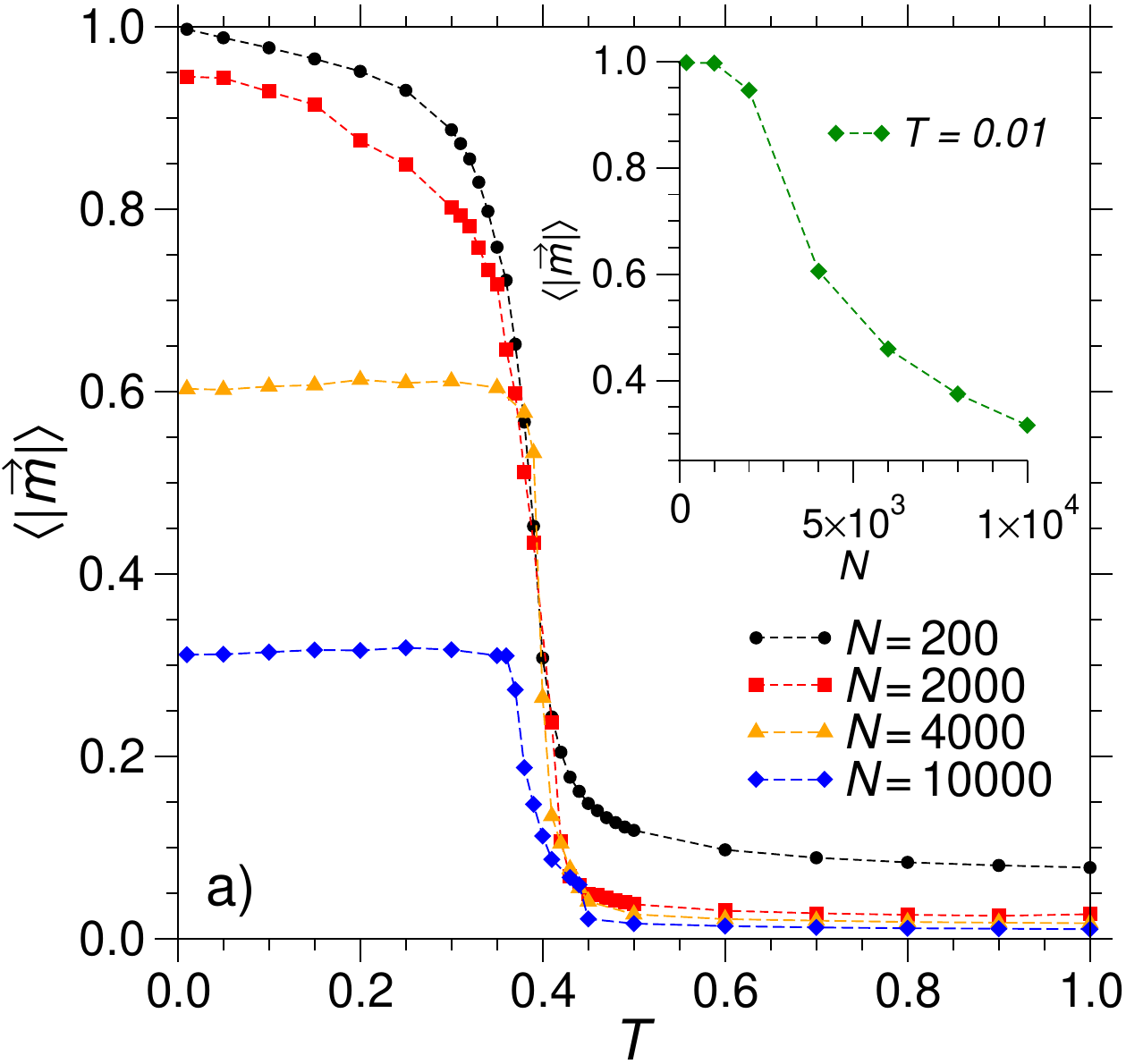}
\includegraphics[width=0.5\textwidth]{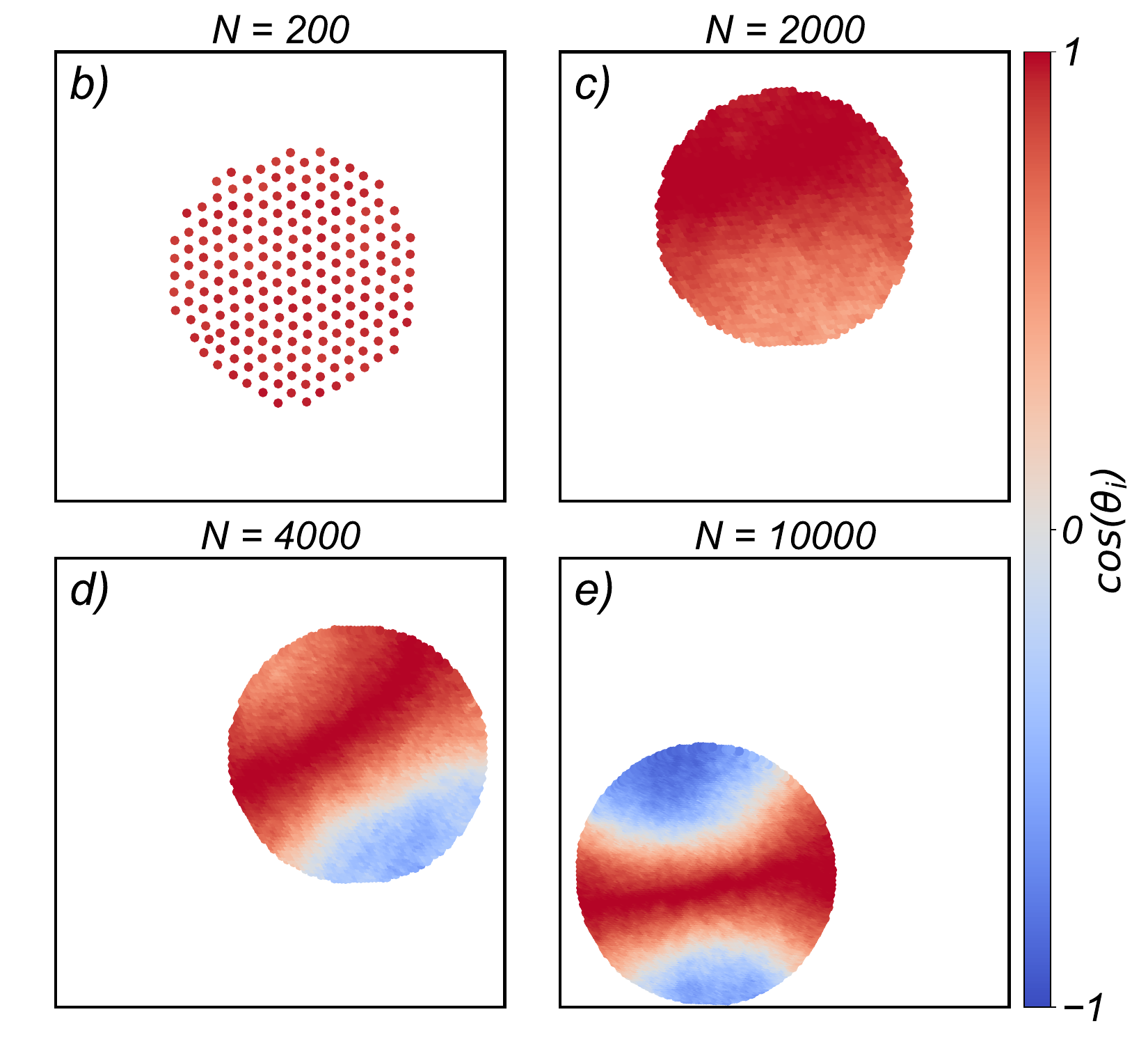}
\caption{a) Average magnetization per particle $\langle|\vec{m}|\rangle$ as a function of temperature for systems with $N=200$, 2000, 4000, and 10000 particles at the density $\rho = 0.2$ and $K=0.1$. The inset shows $\langle|\vec{m}|\rangle$ as a function of $N$ at the fixed temperature $T=0.01$. The snapshots correspond to $T=0.01$, $\rho = 0.2$, and $K=0.1$ for b) $N=200$, c) $N=2000$, d) $N=4000$ and e) $N=10000$ particles. \label{fig9}}
\end{figure*}

{\bf The occurrence of moving clusters.}  Even for temperatures $T< T_c$, properties such as energy, pressure, and specific heat do not seem to depend on the choice of the parameter $K$ (cf.~Figs.~\ref{fig4} and \ref{fig6}). However, as can be inferred from the snapshots in Fig.~\ref{fig5} for $T < T_c$, the average magnetization $\langle | \vec{m} | \rangle$ is strongly affected by the choice of $K$. For a given system size, as we shall see in the following, this quantity decreases with increasing $K$. Furthermore, we will see that in the cluster phase a finite average magnetization is associated with a finite center-of-mass velocity (or group velocity) of the cluster and one observes a motion of the cluster that is reminiscent of collective motion in typical active systems (cf.~for example, with a grain of salt, to flocks of birds).

Figure \ref{fig7} shows distributions of the particle velocities $v_x = \frac{1}{m} (p^x - A^x)$ in $x$ direction, $P(v_x)$, for the high temperature regime where the system is in a gas phase. Of course, in a statistical sense, the same distributions are obtained for the velocities in $y$ direction. To obtain the distributions, we have performed 20 to 400 simulations for $2\times 10^5\,\tau$. As indicated by the black solid lines in Fig.~\ref{fig7}, the distributions $P(v_x)$ (here, at the fixed density $\rho = 0.2$) can be very well described by the Gaussian function,
\begin{equation}
P(v_x) = \sqrt{\frac{m}{2 \pi k_{\mathrm{B}} T}} \, \exp\left( - \frac{m v_x^2}{2 k_{\mathrm{B}} T} \right) \, ,
\label{eq_gaussv}
\end{equation}
corresponding to a zero mean and standard deviation $\sqrt{k_{\mathrm{B}} T}$. For a fixed temperature in the high-temperature regime, $T=0.4$, identical distributions $P(v_x)$ are obtained over a broad range of densities, $0.1 \le \rho \le 0.5$ (see Fig.~\ref{fig7}b). Note that for all the latter densities the system remains in a gas phase.

The velocity distributions completely change for temperatures $T<T_c$ when the system forms a cluster. Now we compute the distribution of the center-of-mass velocity $v_x^{\mathrm{CM}}$ of the system in $x$ direction, $P(v_x^{\mathrm{CM}})$. This distribution is displayed in Fig.~\ref{fig8}a at different temperatures for systems with $N=2000$ particles. Note that again identical results are obtained for the corresponding velocity distribution in $y$ direction. At $T=0.5$, i.e.~above $T_c$, we find a unimodal Gaussian distribution, similar to those shown in Fig.~\ref{fig7}. Around $T_c$, however, at a temperature of $T=0.4$ the distribution becomes bimodal which means that there is the occurrence of a net motion of the center of mass either in positive or negative directions along both axes. This center-of-mass net motion can be interpreted as a flocking motion of the cluster. With decreasing temperature, the two peaks of the bimodal distribution move to higher absolute values of $v_x^{\mathrm{CM}}$. To quantify the average velocity in positive and negative direction, we compute first moments of $P(v_x^{\mathrm{CM}})$ in the following manner:
\begin{eqnarray}
    V_+^{\mathrm{CM}} & = & 2 \int_0^\infty v_x^{\mathrm{CM}} P(v_x^{\mathrm{CM}}) \, \mathrm{d} v_x^{\mathrm{CM}} \\
    V_-^{\mathrm{CM}} & = & 2 \int_{-\infty}^0 v_x^{\mathrm{CM}} P(v_x^{\mathrm{CM}}) \, \mathrm{d} v_x^{\mathrm{CM}} \, ,   
\end{eqnarray}
where $V_+^{\mathrm{CM}}$ and $V_-^{\mathrm{CM}}$ are the first moments of $P(v_x^{\mathrm{CM}})$ with respect to positive and negative values of $v_x^{\mathrm{CM}}$, respectively. Thus, the average center-of-mass velocity $\langle V_x^{\mathrm{CM}} \rangle$ as a function of temperature has a positive and a negative branch that is given by $V_+^{\mathrm{CM}}(T)$ and $V_-^{\mathrm{CM}}(T)$, respectively. The results for the temperature dependence of $\langle V_x^{\mathrm{CM}} \rangle$ are shown in Fig.~\ref{fig8}b for systems with $N=200$ and $N=2000$ particles. This figure indicates that there is a bifurcation around $T_c$ where towards low temperatures the average centre-of-mass velocity splits into a positive and a negative branch. With increasing system size, the magnitude of the velocity tends to decrease.

\begin{figure*}[t]
\centering
\includegraphics[width=0.45\textwidth]{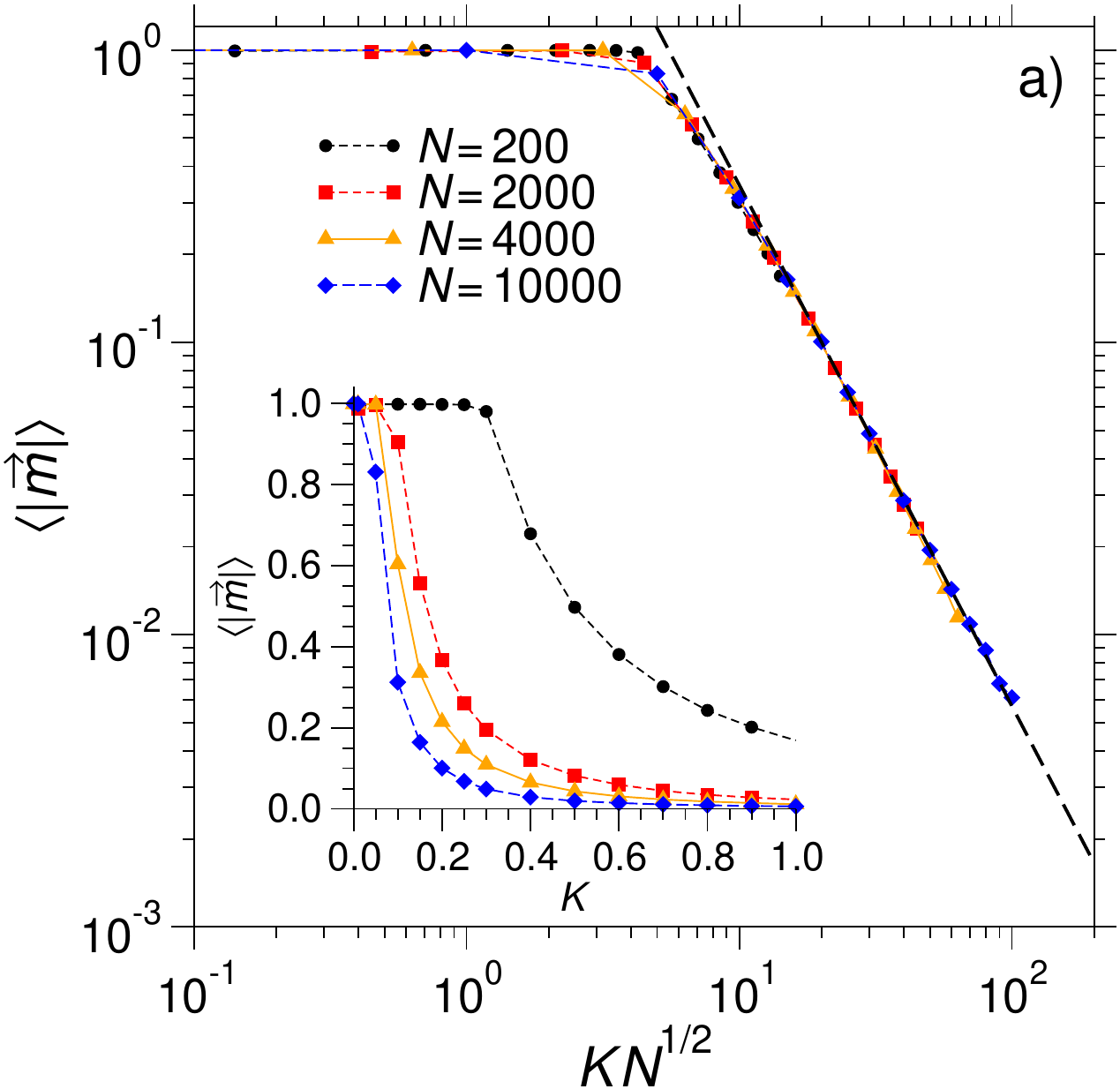}
\includegraphics[width=0.5\textwidth]{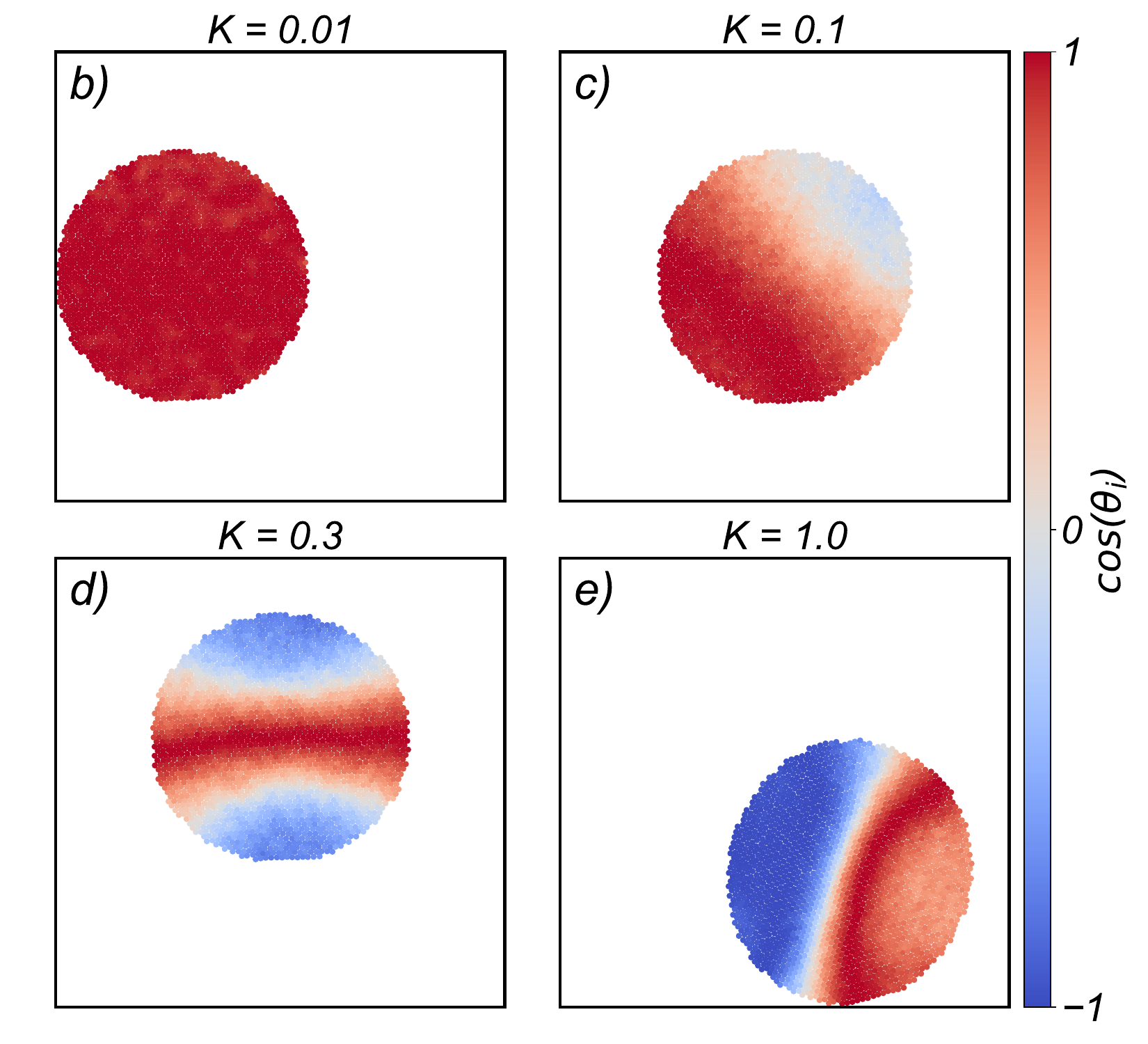}
\caption{a) Double-logarithmic plot of the average magnetization per particle $\langle|\vec{m}|\rangle$ as a function of $KN^{1/2}$. The data points correspond to different system sizes, as indicated, at the density $\rho = 0.2$ and the temperature $T=0.01$. The black dashed line shows the fit function $f(x) = 20.9677 \times x^{-1.782973}$ with $x=KN^{1/2}$ (see text). The inset shows the same data in a linear plot as a function of $K$. The snapshots correspond to $T=0.01$, $\rho = 0.2$, and $N=2000$ particles for b) $K=0.01$, c) $K=0.1$, d) $K=0.3$ and e) $K=1.0$. \label{fig10}}
\end{figure*}

The average center-of-mass velocity is intimately related to the finite average magnetization $\langle | \vec{m} | \rangle$. Since the total canonical momentum of the AH system is conserved, the center-of-mass velocity is given by $\vec{v}^{\, \mathrm{CM}} = - K \vec{m}$. This implies that $\vec{v}^{\, \mathrm{CM}}$ exhibits the same behavior as $\vec{m}$, scaled by a factor of $-K$.

Figure \ref{fig9}a shows the average magnetization $\langle | \vec{m} | \rangle$ as a function of temperature for four systems with $N=200$, 2000, 4000, and 10000 particles at the density $\rho = 0.2$ and $K=0.1$. For all the considered system sizes, the magnetization rapidly decreases towards a low $N$-dependent value around $T_c$. For low temperature $T<T_c$, it has a weak dependence on temperature for the two smaller systems while it is almost a constant for the larger systems. With increasing $N$ this constant decreases and tends to vanish in the limit $N\to \infty$. This can be also inferred from the inset of Fig.~\ref{fig9}a that shows the $N$ dependence of $\langle | \vec{m} | \rangle$ at a fixed temperature $T=0.01$ and coupling constant $K=0.1$. Snapshots for the different system sizes at $T=0.01$ and $K=0.1$ are shown in Figs.~\ref{fig9}b-\ref{fig9}e. These snapshots indicate that for small system sizes the magnetization of clusters is homogeneous while for sufficiently large systems spin patterns emerge that tend to reduce the total magnetization of the cluster. As discussed by Casiulis {\it et al.}~\cite{Casiulis2020_2}, these spin patterns can be interpreted as spin-wave excitations associated with topological defects that emerge because of the high cost of kinetic energy $\sim NK^2$. The polar states with homogeneous magnetization are only observed for small system sizes where collective spin wave excitations with a sufficiently large wave-length and thus a sufficiently low energy are not compatible with the size of the cluster. Thus, there is a competition between the cost of kinetic energy $\sim NK^2$ and the cost of for the spin-wave excitation with the longest wave-length or lowest energy that fits into circular cluster. The crossover from polar to spin-wave states has to depend both on $K$ and $N$.

To elucidate the dependence of the megnetization on the latter parameters, we first consider its dependence on $K$ for different system sizes at constant density $\rho=0.2$ and the low temprature $T=0.01$ (inset of Fig.~\ref{fig10}a). Here, for the smallest system size, $N=200$, we can nicely infer the crossover from the polar state for small $K$ with a constant magnetization close to one to the spin-wave state where  $\langle | \vec{m} | \rangle$ decreases with increasing $K$. With increasing particle number $N$, the $K$ regime of polar states shifts to lower values of $K$, as expected. Note that in the limit $K\to 0$ a magnetization close to one is obtained for all system sizes which indicates the absence of a BKT transition in our system (see also the discussion above). The snapshots, Figs.~\ref{fig10}b to \ref{fig10}e, nicely illustrate the crossover from polar to spin-wave states with increasing $K$ for the example of systems with $N=2000$ particles at $\rho=0.2$ and $T=0.01$.

When plotting the data as a function of $KN^{1/2}$ (main plot in Fig.~\ref{fig10}a), we see that all the data very nicely fall on a single master curve. The crossover from polar to spin-wave states occurs around a value of $x_c \approx 4.25$ (with $x=K N^{1/2}$). This is in nice agreement with an estimate of this crossover point, as proposed by Casiulis {\it et al.}~\cite{Casiulis2020_2}. According to their prediction, adapted to our system, the crossover occurs at $x_c = K_c N_c^{1/2} \approx \frac{1}{2} \pi \bar{r} \sqrt{z \bar{J} \pi \rho_{\mathrm{cl}}}$ with $\rho_{\mathrm{cl}}$ the density of particles in the circular cluster, $\bar{r}$ the average distance between neighboring particle pairs, $z$ the coordination number around the particle, and $\bar{J}$ the average value of the interaction strength of interacting spin pairs. If we set $\bar{r} \approx 1$, $\bar{J} \approx 1$, $z=6$, and $\rho_{\mathrm{cl}} \approx 0.8$, we obtain $x_c \approx 6$ which gives the correct order of magnitude of $x_c$. The decay of $\langle | \vec{m} | \rangle$ for $x>x_c$ can be well described by a power law, $\langle | \vec{m} | \rangle \propto x^{-1.8}$ (black dashed line in Fig.~\ref{fig10}a). We do not have an explanation for the exponent -1.8. 

\section{Conclusion \label{sec_conclusion}}
In this study, we have revealed fundamental issues of two-dimensional AH systems, defined as $XY$ spin fluids with a spin-velocity coupling via a vector potential. Although these systems have non-standard thermodynamic properties, they exhibit most of the symmetry properties of normal Hamiltonian systems in equilibrium. Thus, energy and total canonical momentum is conserved. Moreover, the equipartition theorem in its generalized form \cite{Huang2008} holds, providing a definition of thermal energy and temperature with respect to the kinetic degrees of freedom. Here, temperature cannot be expressed in terms of an average of the kinetic energy as in standard Hamiltonian systems, see Eq.~(\ref{eq_temperature}). Also the pressure, of course, contains additional terms due to the coupling of spins and velocities and is thus not just given by the standard microscopic formula. The simplest version of an AH model is probably the CTCD model that we have simulated in the framework of the present study. At low temperature this model may show states of moving clusters that can be interpreted as active Hamiltonian states, in the sense that such states show typical dynamic patterns of active systems that usually do not occur under equilibrium conditions.

We have derived a symplectic integration scheme for AH models that includes the possible thermostatting in terms of a Nos\'e-Poincar\'e thermostat to perform MD simulations in the canonical ensemble. The latter thermostat draws from Nos\'e's extended Hamiltonian formalism and integrates a Poincar\'e time transformation to evolve the Nos\'e-Poincar\'e equations of motion within a microcanonical ensemble framework. We have validated our symplectic integration scheme by performing MD simulations of the CTCD model in the microcanonical and canonical ensemble. Our method preserves the total energy, linear, and angular momenta of the (extended) Hamiltonian and enables accurate and stable MD simulations over long timescales both in the microcanonical and canonical ensemble. In particular, we have demonstrated that the proposed scheme correctly generates the canonical ensemble for the system's Hamiltonian, as evidenced by quantitative agreement between the specific heat calculated in the canonical and microcanonical ensembles.

In this work, we have investigated ``active phenomena'' in the framework of the CTCD model. At low temperature, this model shows a transition to a dynamical cluster phase where a cluster moves with a finite center-of-mass velocity. As pointed out by Cavagna {\it et al.}~\cite{Casiulis2019reply}, this cluster phase is not a good model for flocks of birds. However, it demonstrates that the spin-velocity coupling may lead to phases with a collective motion of particles. We have shown that the average modulus of the magnetization and thus the cluster's center-of-mass velocity follows a perfect scaling with the variable $KN^{1/2}$ (cf.~Fig.~\ref{fig10}a). This indicates that the CTCD model supports the center-of-mass motion of finite clusters that do not move in the limit $KN^{1/2} \to \infty$ (note that in this limit the magnetization is zero and one obtains, as shown by Casiulis {\it et al.}~\cite{Casiulis2020_2}, a solitonic spin pattern). In a forthcoming work, we present an AH model that also includes other spin-velocity couplings, leading to a rotational collective motion in addition to the translational one of the CTCD model. In fact, this model shows collective dynamical clusters that are more reminiscent to flocks of birds than those of the CTCD model.

One may argue that the idea of an active Hamiltonian system is somewhat misleading since in principle active systems are non-equilibrium systems. However, AH models share features with real active systems such as the collective motion of particle clusters. In this sense, with respect to phenomena and probably also the dynamical phase behavior, AH systems are the equilibrium counterpart of active systems. Unlike non-equilibrium active systems, the thermodynamics of AH systems is well defined, albeit yet unexplored. Thus, by elucidating the thermodynamics of AH models, one may also shed light on the validity of the phenomenological approaches towards a thermodynamics of active matter \cite{Takatori2014, Omar2020, Speck2016, Solon2018, Klamser2018}.

\begin{acknowledgements}
We want to thank Andrea Cavagna, Olivier Dauchot, Matthias Fuchs, Muhittin Mungan, Ignacio Pagonabarraga, and Peter Virnau for useful discussions. We are grateful to the Alexander von Humboldt foundation for financial support.
SK would like to acknowledge the Alexander von Humboldt fellowship for experienced researchers for generous support during SK's visits to Heinrich Heine University of D\"usseldorf. SK also acknowledges funding by intramural funds at TIFR Hyderabad from the Department of Atomic Energy (DAE) under Project Identification No. RTI 4007, the generous support from the Science and Engineering Research Board (SERB) via Swarna Jayanti Fellowship grants DST/SJF/PSA-01/2018-19 and SB/SFJ/2019-20/05, and the research support from MATRICES Grant MTR/2023/000079 from SERB.
\end{acknowledgements}

\bibliography{bibfile}

\end{document}